\documentclass[twocolumn,showpacs,amsmath,amssymb,prb]{revtex4}

\usepackage{amsmath}

\usepackage{amssymb}

\usepackage{graphicx}
\usepackage{dcolumn}

\usepackage{bm}

\usepackage{color}
\usepackage{amsmath}
\usepackage{amssymb}
\usepackage{graphicx}
\usepackage{dcolumn}
\usepackage{bm}
\def\be{\begin{eqnarray}}
\def\ee{\end{eqnarray}}
\def\ba{\begin{array}}
\def\ea{\end{array}}
\def\nn{\nonumber}
\begin{document}

\title{Local density of states of two-dimensional electron systems
  under strong in-plane electric and perpendicular magnetic fields
} 
\author{S. Erden Gulebaglan and I. Sokmen}
\affiliation{Dokuz Eylul University, Physics Department, Tinaztepe
  Campus, 35160 Izmir, Turkey} 
\author{A. Siddiki}
\affiliation{Istanbul University, Physics Department, Beyazit Campus,
  Istanbul, Turkey} 
\affiliation{Harvard University, Physics Department, Cambridge 02138 MA, USA}
\author{R. R. Gerhardts}
\affiliation{Max-Planck-Institut f\"ur Festk\"orperforschung,
  Heisenbergstrasse 1, D-70569 Stuttgart, Germany} 
\pacs{73.43.Cd}

\begin{abstract}
We calculate the local density of states  of a two-dimensional electron
system under strong crossed magnetic and electric fields. 
We assume a strong perpendicular magnetic field which, in the absence
of in-plane electric fields and collision broadening effects, leads to
Landau quantization and the well-known singular Landau density of states.
Unidirectional in-plane electric fields lead to  a broadening 
of the delta-function-singularities of the Landau density of states. This
results in position-dependent peaks of finite height and width,
which can be expressed in terms of the energy eigenfunctions. These peaks
become wider with increasing strength of the electric field and may
eventually overlap, which indicates the onset of inter-Landau-level
scattering, if electron-impurity scattering is considered.
We present analytical results for two simple models and discuss their
possible relevance for the breakdown of the integer quantized Hall
effect. In addition, we consider a more realistic model for an incompressible
stripe separating two compressible regions, in which nearly perfect screening
pins adjacent Landau levels to the electrochemical potential. We also discuss
the effect of an imposed current on the local density of states in the stripe
region. 
\end{abstract}

\maketitle

\section{Introduction}

The integer quantum Hall effect~\cite{vKlitzing80:494} (IQHE) is one
of the most important discoveries of condensed matter physics, 
observed on two-dimensional electron systems (2DES) subjected to
a strong magnetic field $B$ perpendicular to the plane of the system. 
Measuring the low-temperature resistance in such systems, one finds that in
certain $B$-intervals, the ``plateau regimes'' of the IQHE, the longitudinal
resistance vanishes, indicating dissipationless transport, while the Hall
resistance assumes quantized values, $R_H=h/(\nu e^2)$, where $h=2\pi\hbar$ is 
Planck's constant, $e$ the elementary charge, and $\nu$ a positive integer.
An idealized homogeneous 2DES in a heterostructure (like GaAs/AlGaAs)
without any scattering or interaction of the electrons, 
subjected to a homogeneous in-plane electric field, would yield the same
resistance values. For this ideal 2DES,  $\nu=2\pi \ell^2  n_{\rm el}$ is the
filling factor of the Landau levels, with energy eigenvalues
$E_n=\hbar\omega_c(n+1/2)$ for $n=0,\,1,\dots$, $ n_{\rm el}$ is the
electron density, $\ell=\sqrt{c\hbar/e|B|}$ the magnetic length (we use CGS
units, with $c$ the velocity of light), and $\omega_c=e|B|/( m^* c)$ the
cyclotron frequency, with $m^*$ the effective mass ($m^*=0.067m_e$, for GaAs).
In this paper we neglect spin-splitting of the Landau levels and take account
of the spin degree of freedom by a degeneracy factor $g_s=2$. The fact, that
in the IQHE one observes integer values of $\nu$, suggests that, in the
plateau regime, the electrons which carry the current  fully occupy an integer
number of Landau levels, and thus are in states which are energetically
separated from empty states, so that at low temperatures dissipative
scattering processes are suppressed. Then one has to
understand, why the relevant filling factor $\nu$ remains constant over a
$B$-interval of finite width, and which processes limit the plateau regimes
and lead back to dissipative transport. 

If the effective filling factor has to be constant over an interval of
$B$-values, obviously the relevant electron density must change. In the early
attempts to explain this phenomenon, it has been assumed that the Coulomb
interaction between the electrons is unimportant for the understanding of the
IQHE. Localization
theories \cite{PrangeGirvin,Kramer03:172} stated that electron-impurity
interaction leads to localized states in the tails of collision-broadened
Landau levels, which do not contribute to the current transport.  The
existence of such inert localized states may explain the resistance
quantization, even if a homogeneous current distribution is assumed. Such
theories have, however, problems with explaining the enormous accuracy of
better than $10^{-8}$, with which the quantized resistance values can be
reproduced, \cite{Bachmair03:14}
 even in narrow Hall bars with a width of a few
 micrometers. \cite{Siddiki09:17007}  
The confinement of the 2DES to the interior of such Hall bars leads to edge
states in the gaps between bulk Landau levels, which also have been considered
to be important for the transport in the plateau regime of the
IQHE. \cite{Halperin82:2185,Buettiker86:1761} This picture, which leads to
extremely high current densities in the edge states and strongly $B$-dependent
electron density profiles, has been critizised \cite{Chklovskii92:4026}
 because it neglects important screening effects.

The Landau quantization, which alters the energy-independent density of states
(DOS) of the 2DES at $B=0$,
$D_0=m^*/(\pi \hbar^2)$,  to the Landau DOS with sharp peaks around the Landau
energies $E_n$, leads to peculiar screening effects, with nearly perfect
screening if the Fermi energy $E_F$ (or the chemical potential) coincides with a
Landau energy, and with no screening, if $E_F$ falls into a gap between two
Landau levels.\cite{Efros88:1281,Efros88:1019}
In an inhomogeneous 2DES, as e.g. in a Hall bar with electron depletion near
the edges, this should lead to the occurrence of ``compressible regions'',
where a Landau level is pinned to the Fermi energy and the density is
position-dependent, and ``incompressible regions'' which separate neighboring
compressible regions with adjacent Landau levels at $E_F$.  In these
incompressible regions $E_F$ lies in the gap between two Landau levels, and
one expects there a constant electron density corresponding to an integer
value of the filling
factor. \cite{Chklovskii92:4026,Chklovskii93:12605,Lier94:7757,Oh97:13519} 
Scanning force microscope experiments
\cite{Weitz00:247,Ahlswede01:562,Ahlswede02:165} on narrow Hall bars have
confirmed this picture. In the plateau regimes of the IQHE, the Hall potential
across the sample drops across stripes at the  positions predicted for
incompressible stripes (ISs), and is constant elsewhere. Well outside the
plateau regimes, the Hall potential varies nearly linearly between the sample
edges.  Self-consistent calculations for equilibrium
\cite{Lier94:7757,Oh97:13519} 
and transport \cite{Guven03:115327,Siddiki04:195335,Gerhardts08:378} have
clarified the situation. At sufficiently high temperature the conductivity is
Drude-like, the current density $j_y(x)$ along the sample
is proportional to the electron density  $n_{\rm el}(x)$, and the Hall
potential varies linearly across the sample. With decreasing temperature, near
the lines of constant local filling factor  $\nu(x)=2\pi \ell^2 n_{\rm el}(x)
=k g_s$ with even-inter values (since spin-degeneracy is assumed) local minima
of the longitudinal resistivity develop (Shubnikov-deHaas effect), which lead
to local maxima of the current density.  With further decreasing temperature,
ISs with extremely small values of the longitudinal resistivity develop along
these lines and the current becomes confined to these stripes, where it can
flow nearly without dissipation. It has been argued
\cite{Siddiki04:195335,Gerhardts08:378} that an IS supporting dissipationless
transport between two compressible regions can develop only if the distance
between these regions is sufficiently large, larger than several times the
extent of a typical wavefunction, say about $7\ell$, which requires a
sufficiently slow variation of the effective potential. \cite{Siddiki10:113011}
 For the considered class
of samples this guarantees that, in agreement with the experiments,  the
plateau regimes of the IQHE are well separated on the $B$-axis, and that only
ISs with the same value of the local filling factor can exist simultaneously
in the Hall bar.

If the stripe between two compressible regions becomes very small, one may
expect that energy eigenfunctions with centers in different regions overlap and
lead to  quasi-elastic inter-Landau-level scattering (QUILLS), a mechanism
which has been discussed for a long time as a possible reason for the
breakdown of the IQHE under strong imposed
currents. \cite{Eaves86:346,Guven02:155316} Such an overlap of wavefunctions,
belonging to different Landau quantum numbers and having different center
coordinates, but having the same energy eigenvalues, can occur in Landau
levels tilted by a constant (Hall) electric field.  
A suitable quantity to study such overlap effects seems to be the local
density of states (LDOS), which has recently been studied for such a
constant-electric-field model and interpreted with respect to the
IQHE, \cite{Kramer04:21,Kramer06:1243}  however with a rather complicated
mathematical approach and dubious results for the IQHE.

The purpose of this paper is to present a simple way to calculate the
LDOS for a 2DES with translation symmetry in one in-plane direction,
sect.\ref{sec:sec1}, to give explicit analytic and numeric results for simple
model potentials in the other in-plane direction, sect.\ref{sec:exact-mod},
and, finally, to evaluate the LDOS for a simple but realistic model of an IS
between two compressible regions, which carries an intrinsic and, possibly, an
imposed external current, sect.\ref{sec:is-model}.

\section{Electric-field-broadened Landau levels\label{sec:sec1}} 

We describe a 2DES in the $x$-$y$-plane, subjected to a strong magnetic
field $\mathbf{B}=(0,0,B)=\nabla \times \mathbf{A(r)}$ in $z$-direction, in an
effective-field (e.g. Hartree) approximation by a single-particle Hamiltonian 
\be H=\frac{1}{2 m^*}\Big(\mathbf{p}+\frac{e}{c} \mathbf{A(r)}\Big)^2
+V(\mathbf{r}),
\label{eq:hamilton} \ee
where the potential energy $V(\mathbf{r})$ may contain the effect of
externally applied static electric fields, of lateral confinement, and of the
average Coulomb interaction with the other electrons of the 2DES. Once the
eigen-functions $\psi_{\alpha}(\mathbf{r})$ of the Schr\"odinger equation
\be (H-E_{\alpha})\psi_{\alpha}(\mathbf{r})=0 \label{eq:schrodinger} \ee
are known, one can calculate the electron density
\be n(\mathbf{r})= \sum_{\alpha} f_{\alpha} |\psi_{\alpha}(\mathbf{r})|^2,
\label{eq:density} \ee
where the occupation probability  $f_{\alpha}$ of the energy eigenstate
$|\alpha \rangle$ may depend on all the quantum numbers of conserved
quantities collected in $\alpha$, i.e., two for orbital motion and one for
spin. 

\subsection{Local density of states (LDOS)  \label{sec:ldos}}  

If in Eq.~(\ref{eq:density}) the occupation probability of the state $|\alpha
\rangle$ depends only on its energy eigenvalue, $f_{\alpha}=f(E_{\alpha})$, it
may be useful to express the density
\be   n(\mathbf{r})= \int dE\, f(E) \, D(E;\mathbf{r}) \label{eq:denldos}
\ee
in terms of the ``local density of states'' (LDOS):
\be D(E;\mathbf{r})= \sum_{\alpha} \delta(E- E_{\alpha})
|\psi_{\alpha}(\mathbf{r})|^2. \label{eq:ldos1} \ee
This formula for the LDOS is easily generalized to include the effect of
quasi-elastic scattering of the electrons by randomly distributed impurities,
which leads to a ``collision broadening'' of the $\delta$-function in
Eq.~(\ref{eq:ldos1}). Systematic calculations of collision broadening 
\cite{Scher66:598,Keiter67:215,Bangert68:177,Gerhardts75:327} usually
start from the Green operator $G_{\rm imp}(z)=(z-H-V_{\rm imp})^{-1}$ for  a
fixed impurity configuration, 
described by an impurity potential $V_{\rm imp}(\mathbf{r})$, and calculate
approximately the average over all possible impurity configurations. This 
average is  expressed in terms of a self-energy operator $\Sigma(z)$,
\be G(z)=[z-H-\Sigma(z)]^{-1} = \Big\langle G_{\rm imp}(z) \Big\rangle_{\rm
  imp}. \label{eq:greenf} \ee
With $G^+(E)=G(E+i0^+)$ the corresponding generalization of
Eq.~(\ref{eq:ldos1}) reads
\be  D(E;\mathbf{r})=-\frac{1}{\pi} {\rm Im}\langle \mathbf{r} |G^+(E)
|\mathbf{r} \rangle. \label{ldosgreen} \ee
Without impurities $\Sigma(z) \equiv 0$, and Eq.~(\ref{ldosgreen}) reduces to
Eq.~(\ref{eq:ldos1}), with
the notation $\langle \mathbf{r}|\alpha \rangle = \psi_{\alpha}(\mathbf{r})$.  

\subsection{Translation symmetry in $y$-direction 
\label{sec:transl}} 
In the following we assume that the  system is
 translation-invariant in $y$-direction, but electric fields in
$x$-direction, $\mathbf{E}=(E_x,0,0)= \nabla V(x)/e$, will be allowed. The translation
invariance in  $y$-direction suggests the Landau gauge
 $\textbf{A}(\textbf{r})=(0,xB,0)$
for the vector potential, so that the single-electron Hamiltonian
(\ref{eq:hamilton}) 
becomes cyclic in $y$ and allows  the separation ansatz 
\be \psi(x,y)=\frac{e^{iky} }{\sqrt{L_y}}\, \varphi_k(x), \label{eq:ansatz}\ee
where $L_y$ ($\rightarrow\infty$) is the normalization length in $y$
direction, and the quasi-continuous momentum 
quantum number $k$ assumes the values $k=2\pi n_y/L_y$, for
arbitrary integers $n_y$. With this ansatz the Schr\"odinger equation
(\ref{eq:schrodinger}) reduces to the one-dimensional form
\be H_X\varphi_{n,X}(x)=E_n(X)\varphi_{n,X}(x), \ee
with the 
 effective Hamiltonian
\be
H_X=-\frac{\hbar^2}{2m^*}\frac{d^2}{dx^2}+\frac{m^*}{2}\omega_c^2(x-X)^2
+V(x), \label{eq:effH}\ee
where  $X=-\ell^2k$ denotes the center of the parabolic potential,
which describes the effect of the magnetic field and leads for fixed
$X$ to a discrete energy spectrum $E_n(X)$. Here and in the following
we neglect spin splitting and consider spin by a degeneracy factor $g_s=2$.
In general the eigenstates $\langle \mathbf{r}|n,X\rangle$ carry current in
$y$-direction, and the expectation value of the velocity operator $\hat{v}_y$
is given by (Hellmann-Feynman theorem)
\be \langle n,X| \hat{v}_y|n,X\rangle =-\frac{1}{m^*\omega_c}
\frac{dE_n(X)}{dX}\quad \Big[\equiv \frac{1}{\hbar}\frac{dE_n}{d
  k}\Big].  \label{eq:vy} \ee   
Then the electron density, Eq.~(\ref{eq:density}), depends only on $x$,
\be n(x)=
\frac{g_s}{2 \pi \ell^2} \sum_n \int dX \, f_{n,X}\, |\varphi_{n,X}(x)|^2,
\label{eq:dens-fnX} \ee
and is accompanied by a current density
\be j_y(x)=\frac{g_s e}{2 \pi \hbar} \sum_n \int dX \, f_{n,X}\,
\frac{dE_n(X)}{dX} |\varphi_{n,X}(x)|^2. 
\label{eq:curdens-fnX} \ee
The LDOS, Eq.~(\ref{eq:ldos1}), reduces to
\be D(E;x)\! =\!
\frac{g_s}{2 \pi \ell^2} \sum_n\! \int\! dX \delta(E\!-\!E_n(X))\,
|\varphi_{n,X}(x)|^2. \label{eq:ldos2} \ee
If the dependence of $E_n(X)$ on $X$ is smooth enough to allow for a
Taylor expansion around the center coordinate $X_{n,E}$ defined by
$E_n(X_{n,E})=E$, the $X$-integral in Eq.~(\ref{eq:ldos2}) can be
evaluated:
 \be
 D(E;x)=  \frac{g_s}{2 \pi \ell^2} \sum_n \frac{ |\varphi_{n,X_{n,E}}(x)|^2
}{|E'_n(X_{n,E})|}
  \label{eq:ldos-expl}, \ee
%
 with  $E'_n(X_{n,E})=dE_n/dX(X_{n,E})$.
Before we illustrate some properties of this LDOS with typical
examples, we introduce a simple treatment of collision broadening.

\subsection{Collision broadening \label{sec:collbroad}}

\subsubsection{Homogeneous 2DES without electric
  field  \label{sec:ldos1}}  
For $V(x) \equiv 0$ we get the well known Landau problem with 
energy eigenvalues and eigenfunctions
\be
E_n=\hbar\omega_c(n+\frac{1}{2}),\quad\varphi_{n,X}(x)=\frac{1}{\sqrt{\ell}}
\,u_n\Big(\frac{x-X}{\ell}\Big),\label{eq:enpn}\ee 
respectively, where the normalized oscillator wavefunctions,
\be
u_n(\zeta)=\Big(\frac{1}{2^nn!\sqrt{\pi}}\Big)^{1/2}\,H_n(\zeta)
\, e^{-\zeta^2/2}, \label{eq:un}\ee 
are given by the Hermite polynomials $H_n(\zeta)$ of order $n$.\cite{Abramowitz}
Since here the energy eigenvalues are independent of $X$, the
$X$-integral in Eq.~(\ref{eq:ldos2}) reduces to the normalization integral of
the eigenfunctions, and the LDOS reduces to the well known 
Landau DOS of the homogeneous system
\be D(E;x)
=\frac{g_s}{2\pi\ell^2}\sum_n\delta(E-E_n), \label{eq:ldos-delta}\ee 
which does not depend on the position $x$.
To include the effect of collision broadening, one has to evaluate the
self-energy operator. With weak assumptions (like rotation symmetry) on the
impurity potentials, one can show that $\Sigma(z)$ and the Green operator
$G(z)$  are diagonal in the Landau representation, and that the matrix
elements together with the eigen-energies $E_n(X)$ do not depend on
$X$. \cite{Scher66:598,Keiter67:215,Bangert68:177,Gerhardts75:327}  
Then in Eq.~(\ref{eq:ldos-delta}) the singular $\delta(E-E_n)$ is replaced by
a spectral function $A_n(E-E_n)$ of finite width. Depending on the approximation
scheme, several analytical forms for the spectral function have been
obtained. The self-consistent Born approximation
(SCBA)\cite{Gerhardts75:327,Ando82:437} leads, if scattering between different
Landau levels is neglegted,  to a semi-elliptical form,
\be A_n^{\rm SCBA}(E-E_n)=\frac{1}{\pi \Gamma_n} \Big(1-\Big[\frac{E-E_n}{2
  \Gamma_n}\Big]^2\Big)^{\frac{1}{2}},\, \label{eq:Ascba} \ee
while other approaches yield a Gaussian form, \cite{Gerhardts75:285}
\be A_n^G(E-E_n)=\frac{1}{\sqrt{2 \pi}\, \Gamma_n} \exp\Big(-\frac{1}{2}
\Big[\frac{E-E_n}{\Gamma_n}\Big]^2\Big). \label{eq:Agauss} \ee 
In the limit of short-range impurity potentials the matrix elements of the
self-energy and thereby the $\Gamma_n$ in Eqs.~(\ref{eq:Ascba}) and
(\ref{eq:Agauss}) become even independent of the Landau quantum number $n$.

\subsubsection{Model for non-homogeneous 
systems \label{sec:phencolbr}}

We now consider the more general case that the effective Hamiltonian
(\ref{eq:effH}) contains a position-dependent potential $V(x)$. For each value
$X$ of the center coordinate this leads again to a discrete energy
spectrum, but these generalized Landau energies usually depend on $X$ and form
dispersive Landau bands with energies $E_n(X)$.

As mentioned above, in the absence of electric fields the Landau energy
eigenvalues do not depend on the center coordinate $X$, and as a consequence
Green and self-energy operator are diagonal in the Landau representation and
independent of $X$. In the presence of electric fields, however, the
eigen-energies depend on $X$, and so do Green and self-energy
operators. Moreover we can show that both are no longer diagonal in the
generalized Landau representation, which diagonalizes the Hamiltonian in the
absence of collisions. This more complicated situation is found even in the
case of  short-range impurity potentials, which for homogeneous systems with
$V(x)\equiv 0$ leads to the same collision broadening for all Landau
levels. Then the 
evaluation of collision broadening effects becomes much more complicated, and
exceeds the scope of the present paper.

For weak electric field $E_x$, on the other hand, the formula
\be D(E;x)\!=\! \frac{g_s}{2 \pi \ell^2}\!\! \sum_n\!\! \int\!\! dX\,
A_n\big(E\!-\!E_n(X)\big) \,
\varphi_{n,X}^2(x),\,~ \label{eq:plin-cb} \ee
which is correct for $E_x=0$ and a straighforward generalization of
Eq.~(\ref{eq:ldos1}), should yield a reasonable description of
collision broadening effects.
Since we are not aware of a better, practicable approach to treat collision
broadening in the presence of electric fields, we will in the following use
Eq.~(\ref{eq:plin-cb}) as a phenomenological rule to estimate the consequences
of such scattering effects.

\section{Exactly solvable models \label{sec:exact-mod}}

\subsection{Constant electric field\label{sec:ldos2}}
Simple analytic results are also obtained for the case of a constant
in-plane electric field $\mathbf{E}=(E_x,0,0)$, leading to the
potential $V(x)=e x E_x$. Within classical mechanics, this leads for
an ideal 2DES to a constant Hall drift of the centers of the cyclotron
motion, which can be eliminated by a Galilei transformation to a
coordinate system moving with the drift velocity $\mathbf{v}_D= c
\mathbf{E \times B} /B^2=(0,-cE_x/B,0) $. Since all electrons suffer the same drift velocity,
the current density $\mathbf{j}(x)=-e \mathbf{v}_D n(x)$ is proportional to
the electron density $n(x)$, and one obtains Ohm's law $\mathbf{j}(x)=
\hat{\sigma}(x) \mathbf{E}$ with the Hall conductivity
$\sigma_{yx}(x)=(ec/B)n(x)$ and vanishing longitudinal conductivity,
$\sigma_{xx}(x) \equiv 0$. 

\subsubsection{Eigenstates and LDOS}
Inserting  $V(x)=e x E_x$ into the Hamiltonian (\ref{eq:effH}) results in a
shifted parabolic potential with the new center
$\tilde{X}=X-eE_x/(m^*\omega_c^2)$ and position-independent terms, which add to
the oscillator energies $\varepsilon_n =\hbar\omega_c(n+1/2)$.
The resulting energy eigenvalues and eigenfunctions
are
\be \tilde{E}_n(\tilde{X})&=&\varepsilon_n+eE_x\tilde{X}+\frac{m^*}{2}v_D^2
\nn \\ 
 &=&\varepsilon_n+eE_x X -\frac{m^*}{2}v_D^2 \equiv E_n(X),\label{eq:enX}\ee 
and
\be  \varphi_{n,X}(x)
=\frac{1}{\sqrt{\ell}} \,
u_n\Big(\frac{x-\tilde{X}}{\ell}\Big),
 \label{eq:fitilde}\ee  
respectively, with $v_D=cE_x/B$.
From Eq.~(\ref{eq:vy}) we see that each state carries the same current 
$-e\langle n,X|\hat{v}_y|n,X\rangle =e^2E_x/m^*\omega_c=ev_D$, in analogy to the
fact, that the radius of the classical cyclotron orbit has no influence on the
drift velocity of its center. As a consequence of Eqs.~(\ref{eq:dens-fnX}) and
(\ref{eq:curdens-fnX})  the current density is
directly proportional to the electron density,
\be j_y(x)=ev_D n(x), \label{eq:jy-class} \ee
independent of the occupation probability of the eigenstates, just as in the
classical case.

Due to the linear dependence of $
\tilde{E}_n(\tilde{X})$ on $\tilde{X}$, Eq.~(\ref{eq:ldos-expl}) can be
written as
\be
D(E;x)=\frac{g_s}{2\pi\ell^2} \sum_n \frac{1}{e|E_x|\ell}
\,u_n^2\Big(\frac{\tilde{E}_n(x)-E}{eE_x\ell}\Big). \label{eq:plinldos}\ee 
\begin{figure}[h]
 \includegraphics[width=0.9\linewidth]{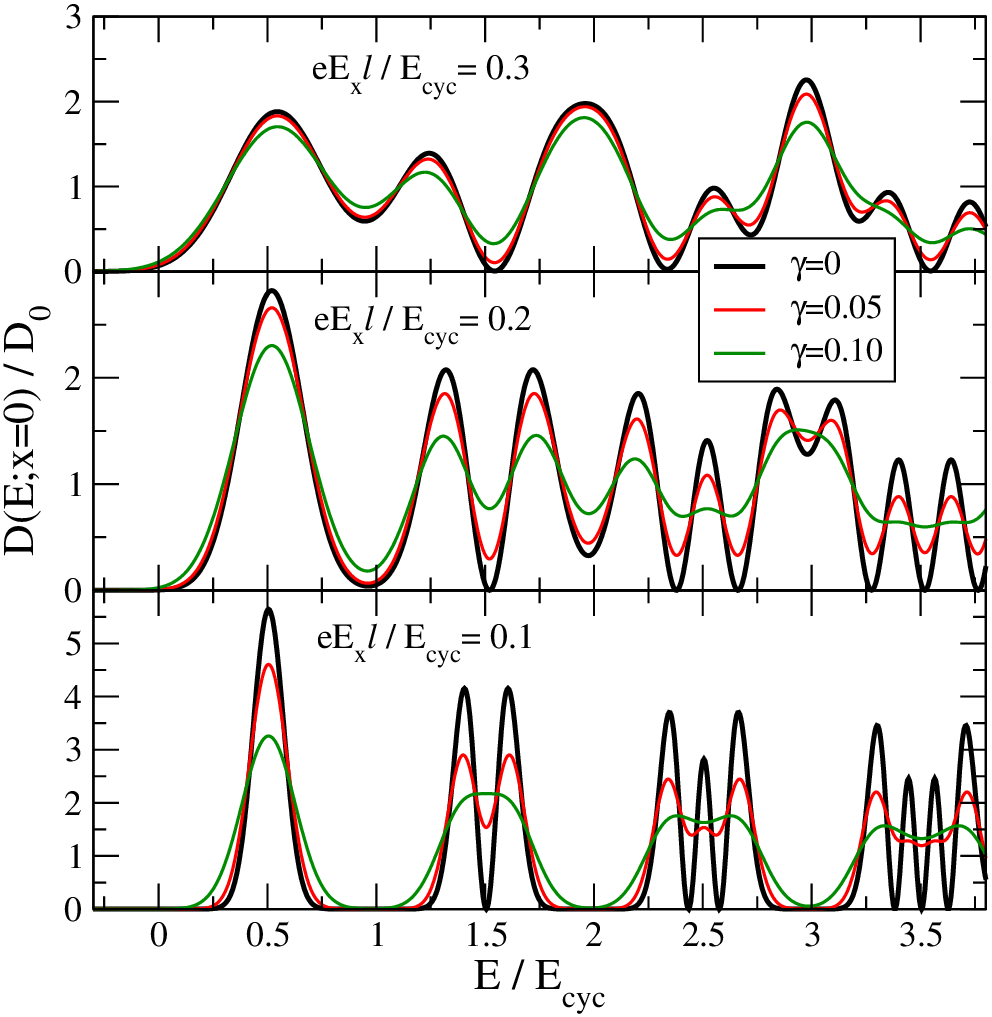}
\caption{\label{fig:plin-ldos} (color online) Heavy black lines ($\gamma=0$):
  LDOS according to 
  Eq.~(\ref{eq:plinldos}) for three values of the electric field strength,
  $|eE_x|\ell/\hbar \omega_c=0.1,\, 0.2$ and $0.3$. Also shown is the effect of
  level broadening according to Eqs.~(\ref{eq:plin-cb}) and (\ref{eq:Agauss})
  for   $\gamma\equiv \Gamma/\hbar 
  \omega_c= 0.05$ and $0.1$. $D_0=m^*/(\pi \hbar^2)$,  $g_s=2$.}
\end{figure}
This result has been obtained in Ref.~[\onlinecite{Kramer04:21}] in a much
less transparent way, starting from the symmetric instead of the Landau gauge
for the vector potential.

Since in Eq.~(\ref{eq:plinldos})  the argument of the wavefunctions
depends linearly on both, the position $x$ and the energy $E$,
the energy  dependence of the LDOS reflects the
position dependence of the energy eigenfunctions. The
contributions of the individual Landau levels, which become $\delta$-function
like for vanishing $E_x$, become wider with increasing electric field, with a
width proportional to $eE_x\ell$. If one shifts the position from $x$ to
$x+a$, one obtains the same profile for the LDOS, but shifted on the energy
axes by $-eE_xa$:  $D(E;x+a)=D(E-eE_xa;x)$.
Typical results \cite{Kramer04:21} for the LDOS  according to
Eq.~(\ref{eq:plinldos}) are shown by the heavy black lines in
Fig.~\ref{fig:plin-ldos}. 
It is seen that the gaps between the lowest adjacent Landau levels close for
$0.1< eE_x\ell/\hbar \omega_c < 0.2$, whereas the zeroes of the LDOS
determined by the zeroes of the eigenfunctions remain rather stable.

If contributions to the LDOS due to adjacent Landau levels overlap,
this means that, at
the same position and at the same energy wavefunctions due to different Landau
levels have finite values. Then, if there is also a non-vanishing
impurity potential at 
this position, the potential matrix element between these adjacent levels is
finite and there must be elastic scattering between these levels.
This quasi-elastic inter-Landau-level scattering (QUILLS) has been discussed
for a long time \cite{Eaves86:346,Guven02:155316}
as a possible  mechanism for the breakdown of the IQHE under strong imposed
currents. Apparently QUILLS must become important in the neighborhood of
narrow incompressible strips, where the local potential must bridge an amount
of order $\hbar \omega_c$ across a strip of a width less than about $10\ell$.

\begin{figure}[h]
 \includegraphics[width=0.9\linewidth]{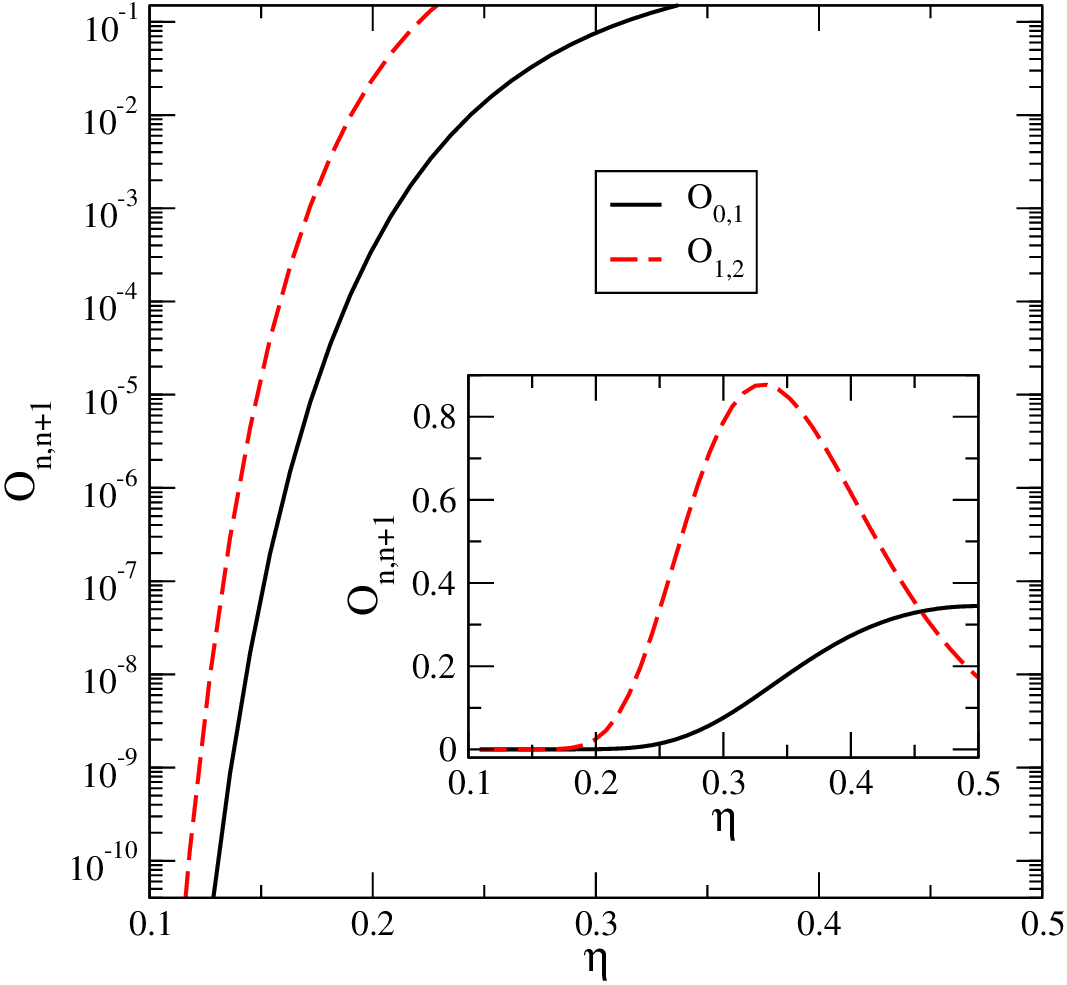}
\caption{\label{fig:overlp} Dimensionless overlap for the two lowest Landau
  gaps as function of $\eta=e|E_x|\ell/\hbar \omega_c$. }
\end{figure}

 The colored lines in Fig.~\ref{fig:plin-ldos}
are calculated  from Eq.~(\ref{eq:plin-cb}) for a Gaussian spectral function,
Eq.~(\ref{eq:Agauss}), with $n$-independent $\Gamma_n=\gamma\, \hbar \omega_c$
for two values of $\gamma$. Apparently the zeroes of the LDOS, which are due to
the zeroes of the energy eigenfunctions, are smeared out already by a very weak
collision broadening, and are of no importance in real samples.
A discussion \cite{Kramer04:21}
of a possible importance  of these zeroes for the QHE
 is therefore without any relevance. 
On the other hand, the value of the LDOS in the gap between two adjacent
Landau levels is of importance. In order to yield a plateau in the IQHE, the
gap in an incompressible strip between two adjacent compressible regions must
be sufficiently well developed. As a measure for the quality of such gaps we may
consider the overlap of the contributions of adjacent Landau levels to the
LDOS, according to Eq.~(\ref{eq:plinldos}). We define the overlap as the
product of these contributions in the middle $E_{n,n+1}(x)=[\tilde{E}_n(x)+
\tilde{E}_{n+1}(x)]/2$ between these levels, devided by the square of the
zero-$B$ DOS $D_0=m^*/(\pi \hbar^2)$, to make the overlap dimensionless. Since
$g_s/(2\pi \ell^2 D_0)= \hbar \omega_c$, Eqs.~(\ref{eq:enX}) and
(\ref{eq:plinldos}) yield for the
dimensionless overlap of level $n$ and $n+1$:
\be O_{n,n+1}(\eta)=\frac{1}{\eta^2}\, u_n^2\big( -\frac{1}{2\eta}\big)\,
u^2_{n+1}\big(\frac{1}{2\eta}\big), \label{eq:overlap} \ee 
with $\eta=e|E_x|\ell/\hbar \omega_c$. The results for the lowest gaps,
$O_{0,1}(\eta) =\exp(-1/2\eta^2)/(2\pi\eta^4)$ and
$O_{1,2}(\eta)=O_{0,1}(\eta)(2-1/\eta^2)^2/8$, are plotted in
Fig.~\ref{fig:overlp}. 
If we say that the gap between Landau level $n$ and $n+1$ is well developed if
$O_{n,n+1}<10^{-8}$, this defines a critical value $\eta_{n,n+1}^{cr}$
($\eta_{0,1}^{cr} \approx 0.15$, $\eta_{1,2}^{cr} \approx 0.13$)
and thereby a critical field-strength
$E_{n,n+1}^{cr}=\eta_{n,n+1}^{cr} \hbar \omega_c/e\ell$. Only for sufficiently
small electric fields with
\be |E_x| \lesssim 2.1 \eta_{n,n+1}^{cr} (B/ 10{\rm T})^{3/2}\times 10^6\,
{\rm V/m} \label{eq:Exlimit} \ee 
the gap between the Landau levels $n$ and $n+1$ is well developed.
An equivalent formula, with $2.1 \eta_{n,n+1}^{cr}$ replaced by
$1/[\sqrt{2n+1}+\sqrt{2n+3}]$, was given in Eq.~(41) of
Ref.~\onlinecite{Kramer04:21} as experimentally verified criterion for the
breakdown of the  IQHE.

\subsubsection{Occupation of eigenstates}
The properties of a 2DES are not only determined by the properties of the
single-particle eigenstates, but also by their occupation. Here we consider
two different examples.

According to the general rules of statistical mechanics, a thermal
equilibrium state is characterized by the expectation values of its conserved
quantities, which in our case are (1) the particle number, or for fixed volume
the average particle density, (2) the energy, and (3) the quasi-momentum in
$y$-direction, $\hat{p}_y$ with eigenvalues $\hbar k=-m^* \omega_c X$, related
to the translation symmetry.  
The grand canonical distribution function under these boundary conditions
yields
\be f_{n,X}=f(\beta[E_n(X)-\mu -\hbar v_y X/\ell^2]), \label{eq:grandcan} \ee
with $f(x)=1/(1+e^x)$ the Fermi-Dirac function and $\beta=1/k_BT$, $-\beta
\mu$, and  $-\beta \hbar v_y /\ell^2$ Lagrange multipliers conjugated to the
conserved quantities energy, particle number, and quasi-momentum, respectively.

\paragraph{Constant electron density.}
If we choose $v_y=v_D$, equal to the classical Hall drift velocity, the
argument of the distribution function becomes independent of $X$, and the
$X$-integral in Eq.~(\ref{eq:dens-fnX}) reduces to the normalization integral
of the wavefunctions, so that the electron density becomes independent of the
position $x$ and is given by
\be n(x)\equiv \bar{n}_{\rm el}
=\frac{g_s}{2\pi \ell^2} \sum_n f(\beta[\varepsilon_n -m^*v_D^2/2
-\mu]), \label{eq:homHall} \ee
which apart from an unimportant shift of the energy zero is the same as in the
absence of the electric field $E_x$. This choice of the Lagrangian
multiplier apparently describes the quantum analog of the homogeneous Hall
system, with spatially constant electron and current densities.

Considered as a function of the chemical potential, the electron density $
\bar{n}_{\rm el}(\mu)$ for fixed magnetic field is a step-function, with steps
of height $g_s/(2\pi\ell^2)$ and a width of the order of $k_BT$ (or of
$\Gamma_n$ if collision broadening is considered), located near the Landau
energies $\mu \approx \varepsilon_n$.
At low temperatures, $k_BT\ll \hbar \omega_c$,
(and for weak collision broadening, $\Gamma_n\ll \hbar \omega_c$) these steps
are separated by 
wide plateaus of constant  $\bar{n}_{\rm el}(\mu)$, where $\mu$ varies in the
gap between two adjacent Landau levels (LLs). In these plateaus there are no
states 
at the Fermi energy (i.e. near $\mu$) and no quasi-elastic scattering is
possible. Therefore, the longitudinal conductivity is zero and the Hall
conductivity has the quantized value $\sigma_H=e^2\nu/h $ with an integer
value of the filling factor $\nu=2\pi \ell^2 \bar{n}_{\rm el}(\mu)$, even if
one allows for quasi-elastic impurity scattering, which may lead to
dissipation if $\mu$ is located in a broadened  LL.
This scenario is considered in Ref.~[\onlinecite{Kramer04:21}] in order to
explain the IQHE.
The same physical situation can be considered for constant $\mu$ and varying
magnetic field $B$. With increasing $B$ both the degeneracy $g_s/(2\pi\ell^2)$
of the LLs and their energies $\varepsilon_n$ increase linearly with
$B$. If $\mu$ is within a temperature and collision broadened LL,
we may observe dissipation and the conductivity components are not quantized.
With increasing $B$ the energy of this broadened level rises above the
chemical potential, which then falls into the gap below this
broadened level. Then the electron density increases at constant filling
factor linearly with $B$, and the conductivity components are quantized, until
the next lower broadened LL reaches the energy $\mu$. The dissipation
sets in again, and the filling factor decreases by $g_s$ as this level passes
the fixed chemical potential, which causes a rapid decrease of the electron
density.  

If in the low-temperature transport experiments
on 2DESs the chemical potential would be constant, this scenario would
explain the IQHE. This scenario, which at low filling
factors requires that changing the magnetic field induces 
large density changes (up to 50\%), is, however, unrealistic. In real
experiments such a large electron exchange between the 2DES and its
surrounding is hardly possible, and the assumption of 
constant electron density is much more realistic than that of constant
chemical potential. For constant $\bar{n}_{\rm el}(\mu)$, Eq.~(\ref{eq:homHall})
requires that $\mu$ oscillates as a function of $B$, and these oscillations
have indeed been observed experimentally. \cite{Wei97:2514,Wei98:496}
 If a broadened LL is
completely occupied, a further lowering of $B$ at constant 
$\bar{n}_{\rm el}(\mu)$ requires, that $\mu$ jumps to the lower edge of the
next higher LL, so that the quantized values of the conductivity components
occur only at a single, isolated value of $B$, not in a whole
$B$-interval. This explains the well-known Shubnikov-deHaas effect, but not
the IQHE as is claimed in Ref.~[\onlinecite{Kramer04:21}]. The attempts of
Refs.~[\onlinecite{Kramer04:21}] and [\onlinecite{Kramer06:1243}] to calculate
the conductivity tensor are also not 
compatible with accepted transport theories
\cite{Gerhardts75:327,Ando82:437} and yield incorrect results. 

If one writes in Eq.~(\ref{eq:grandcan}) $\mu^*(X)=\mu-\hbar v_y X/\ell^2$ and
replaces the center coordinate by the position  $x$, one gets a
position-dependent electrochemical potential  $\mu^*(x)$. This may reasonably
describe a 
stationary non-equilibrium, and possibly dissipative, state as is found in a
quantum Hall system outside the plateau regime of the
QHE.\cite{Guven03:115327,Siddiki04:195335} In a thermal equilibrium state,
however,  the electrochemical potential $\mu^*$ must be spatially constant and
can depend only on eigenvalues of conserved quantities, such as $X$ and
$E_n(X)$. 

A 2DES with homogeneous electron and Hall current densities may be a good
approximation to the interior of a laterally confined system, far away from the
edges. The same single particle states and energies may, however, also be
used to describe an edge region of a laterally confined system.

\paragraph{Variable electron density.} Let us now
focus on an edge region and assume that the confinement potential there can
be approximated by the linear potential $V(x)=eE_xx$. Let us further assume
that the total, laterally confined, system is in thermodynamic equilibrium,
with vanishing total current. To describe such a state, we put in
Eq.~(\ref{eq:grandcan}) $v_y=0$ and take a fixed constant value $\mu^*$ for
the electrochemical potential. 
\begin{figure}[h]
 \includegraphics[width=0.9\linewidth]{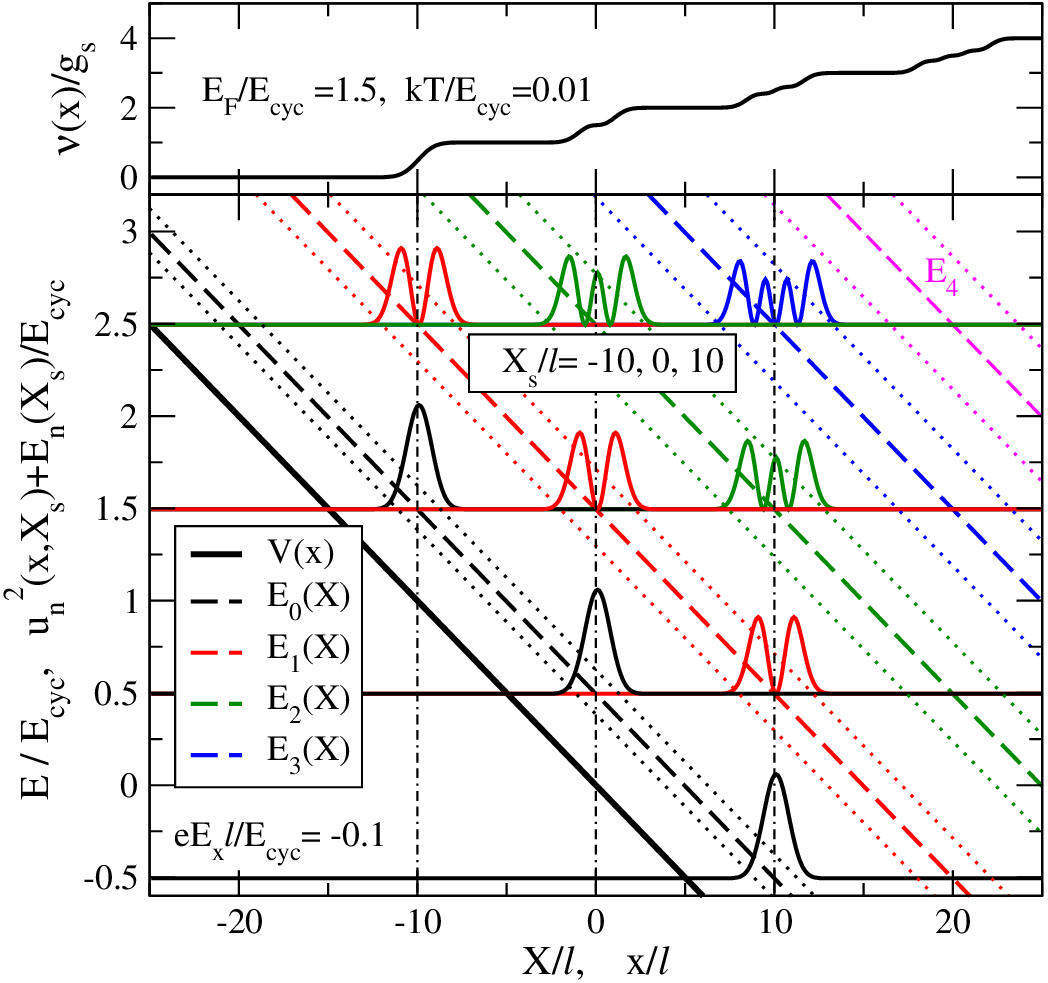}
\caption{\label{fig:plin-wf-dens} (color online) Upper part: filling
  factor $\nu(x)=2\pi \ell^2 n(x)$ for $\mu^*=1.5 \hbar \omega_c$, $k_BT=0.01
  \hbar \omega_c$; lower part: potential $V(x)=eE_xx$, energies $E_n(X)$
  (dashed lines) and the wavefunctions squared at $X_s=0$ and $X_s=\pm
  \hbar \omega_c/eE_x$, 
  shifted by their eigenenergies in units of the cyclotron energy
  $E_{\rm cyc}=\hbar \omega_c$. In both parts $\eta= eE_x\ell/\hbar
  \omega_=-0.1$. The dotted lines indicate the spatial extent of the
  wavefunctions. }  
\end{figure}
Then the electron density can be calculated directly from
Eq.~(\ref{eq:dens-fnX}) or, equivalently, from the LDOS (\ref{eq:plinldos}),
\be n(x)&=&\frac{g_s}{2\pi \ell^2} \sum_n\int \frac{d\tilde{X}}{\ell}
f\Big(\frac{\tilde{E}_n(\tilde{X}) -\mu^*}{k_BT}\Big)\,
u_n^2\Big(\frac{x-\tilde{X}}{\ell}\Big)\nn \\
&=& \int dE \, f\Big(\frac{E -\mu^*}{k_BT}\Big) D(E;x).
\label{eq:dens-plin} \ee

In the lower part of Fig.~\ref{fig:plin-wf-dens} we show, for $E_x<0$, the
linear potential $V(x)=eE_xx$, the lowest energy eigenvalues $E_n(X)$, and,
for $X=0$ the squared eigenfunctions
$u_n^2(x,X)=u_n^2([x-\tilde{X}]/\ell)$, as defined in 
Eq.~(\ref{eq:un}), but shifted upwards by the amount of their energy
eigenvalue. Also shown are the squared eigenfunctions at $X_{\pm}=\pm \hbar
\omega_c/eE_x$, where  $E_n(X_{\pm})=E_{n\pm 1}(0)$. Obviously the distance
$|X_{\pm}|$ 
between the center coordinates of eigenfunctions, which have the same energy
but belong to adjacent Landau levels, is inversely proportional to the field
strength $|E_x|$, whereas the shape of the wavefunctions is independent of
$E_x$. Their  spatial extent, $|x-\tilde{X}| \lesssim 1.2 R_n$, is
indicated by the dotted lines in Fig.~\ref{fig:plin-wf-dens},
 which we estimate by the Landau radius $R_n=\ell \sqrt{2n+1}$ defined
 by $(m^*/2)\,\omega_c^2R_n^2=\hbar \omega_c(n+1/2)$.
Since the distance between the eigenfunctions shrinks with increasing
$|E_x|$, their overlap increases, and is larger for higher than for lower
Landau quantum numbers. This increasing spatial overlap of wavefunctions with
the same energy eigenvalue has, of course, the same origin as the increasing
energetic overlap in the LDOS $D(E;x)$ at a fixed position $x$. 

The upper part of Fig.~\ref{fig:plin-wf-dens} shows the electron density
calculated from Eq.~(\ref{eq:dens-plin}) for $\mu^*=1.5 \hbar \omega_c$, i.e.,
the energy for which the eigenfunctions with $n=0,\, 1$, and $2$ are
indicated. Apparently $n(x)$ increases with $x$ in a stepwise manner, with
steps at positions where wavefunctions of states with energy $E_n(X)\lesssim
\mu^*$ become relevant. The extent of the wavefunctions determines the width of
the steps, and their zeroes lead in the limit $T\rightarrow 0$ to zeroes of
the slope $dn/dx$. This internal structure of the steps occurs on a length of
the order of the magnetic length $\ell$ ($\sim 10$nm for $B \sim 10$T), and
has never been resolved in real Hall bars (with a width $\gtrsim 10\,\mu$m). 
The width of the plateaus between the steps increases
inversely proportional to $|E_x|$. We want to emphasize that a change of the
electrochemical potential $\mu^*$ affects the density profile only by a rigid
shift, $n(x;\mu^*+\delta \mu)=n(x-\delta \mu/eE_x;\mu^*)$.

In the limit  $T\rightarrow 0$ the density is given by 
\be n(x;E_F)=\int_{-\infty}^{E_F} dE \, D(E;x), \label{eq:denst0} \ee
with $E_F=\mu^*(T=0)$. On the scale of Fig.~\ref{fig:plin-wf-dens} this cannot
be distinguished from the given result for $k_BT=0.01 \hbar \omega_c$.  

Of course Eq.~(\ref{eq:jy-class}) yields for the Hall conductivity the trivial
result $\sigma_H(x)=(ec/B)n(x;E_F)$, in agreement with Eq.~(22) of
Ref.~[\onlinecite{Kramer06:1243}] (which considers $j_x$ and $E_y$ instead of
our $j_y$ and $E_x$). To model the current through a macroscopic
device  by $I_y=\int_0^Wdxj_y(x)$ with $j_y(x)=\sigma_H(x)E_x$ calculated from
the present constant-$E_x$ model, as is done in
Ref.~[\onlinecite{Kramer06:1243}], is not meaningful since it effectively   
introduces a very unphysical description of the sample edge at
$x=W$. Describing this edge by a reasonable confinement potential, and the
state of the system by the Fermi-Dirac distribution with a constant Fermi
energy, would lead to vanishing total current, $I_y=0$. To describe a
dissipative state with non-vanishing total current, one needs a
position-dependent electrochemical potential. A dissipation-free state with
finite total current can be described as we explained above, but not with a
position-independent distribution function that depends only on
energy.

In Ref.~[\onlinecite{Kramer04:21}] $n(x;E_F)$ was calculated and discussed for
$x=0$ as function of $E_F$. It was speculated that the structure of this curve
might be related to the quantized Hall effect. This is, however,
incorrect. The QHE is observed on real, confined systems, where a finite total
current, and voltages along and across the sample, can be measured. The
resistance quantization is a property of the sample as a whole. It can
not be explained by local
properties like energy dependence of the electron density at a
position $x$ somewhere inside the sample. The present
linear-potential model, on the other hand, leads, if taken serious, to an 
electron density $n(x;E_F)$ and a current density $j_y(x)=(ecE_x/B) n(x;E_F)$,
which increase with increasing $x$, just because the number of eigenstates
with energy eigenvalues $E_n(X)<E_F$ increases with increasing $X$.
 This model makes
sense only as an approximation to an edge region of a laterally confined
sample, which then, with the choice of a constant electrochemical potential,
has vanishing total current. The property of the $n(x=0;E_F)$-curve tells
nothing about such a real confined sample. 
As is seen from Eq.~(\ref{eq:plinldos}), 
the energy-dependence of the LDOS at fixed position contains exactly the same
information as the position-dependence at fixed energy. Therefore, the
$E_F$-dependence of  $n(x;E_F)$ at fixed $x$ does not contain more information
than the density profile at fixed $E_F$, and does tell nothing about the QHE.

The stepwise increase of the electron density seen in
Fig.~\ref{fig:plin-wf-dens} results, of course, from the smooth linear
increase of the model potential, which is  unrealistic, since it
neglects screening effects, and is
energetically unfavorable in the presence of strong magnetic fields,
as has been emphasized by Chklovskii {\em et al.} \cite{Chklovskii92:4026} and
following work.  

\subsection{Parabolic confinement potential
 \label{sec:parabpot}} 

\begin{figure}[h]
 \includegraphics[width=0.9\linewidth]{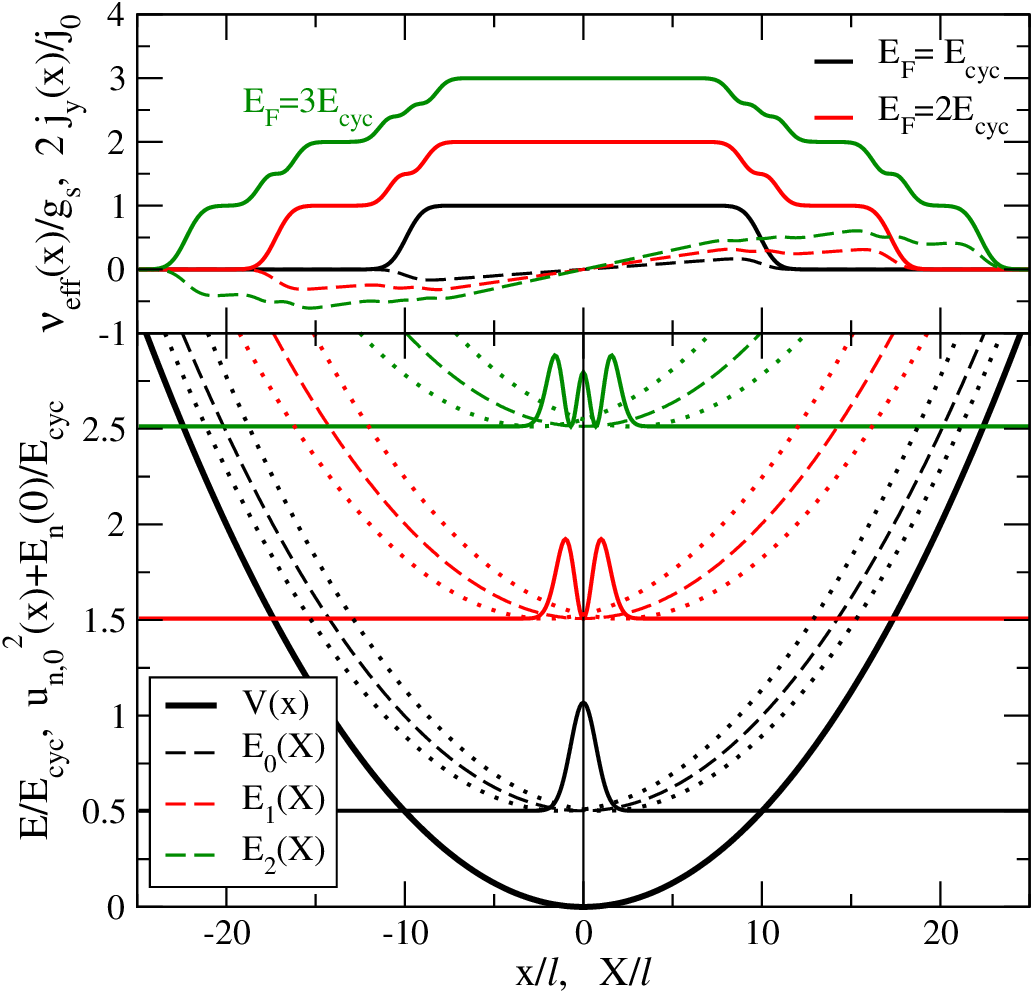}
\caption{\label{fig:parab-wf} (color online) Lower part: potential
  $V(x)=(m^*/2)\Omega^2x^2$ for $\Omega/\omega_c=0.1$, resulting energy bands
  $E_n(X)$ for $n=0,\,1, \,2.$, and squared and shifted eigenfunctions at
  $X=0$. Energies are in units of  the cyclotron energy $E_{\rm cyc}=\hbar
  \omega_c$ and lengths in units of the magnetic length $\ell$. Upper part:
  effective filling factor $\nu_{\rm eff}(x)$
(solid lines) and current density $j_y(x)$ (dashed lines,
  $j_0=g_s e\omega_c /(2\pi \ell)$)
  for three values of the Fermi   energy, see text.
}
\end{figure}

Another simple but instructive model, that also does not describe screening
effects but 
allows to consider closed, laterally confined equilibrium systems and to
calculate the energy eigenvalues and -functions analytically, is  the model of
a parabolic confinement potential, which we 
write as $V(x)=(m^*/2)\, \Omega^2 x^2$. Combined with the parabolic potential
describing the effect of the magnetic field, this leads to the effective
potential
\be && \frac{m^*}{2}\big[ \omega_c^2 (x-X)^2+\Omega^2 x^2\big]\nn \\
&& \hspace*{1.5cm}= \frac{m^*}{2}\big[\widetilde{\omega}^2
(x-\tilde{X})^2+\frac{\omega_c^2}{ 
\widetilde{\omega}^2} \Omega^2 X^2\big], \label{eq:ppeff} \ee
with $\widetilde{\omega}=\sqrt{\omega_c^2+\Omega^2}$ and
$\tilde{X}=(\omega_c/\widetilde{\omega})^2 X$. Energy eigenvalues and
-functions are immediately read off, and can be written as
\be E_n(X)&=&\hbar \widetilde{\omega}\Big[ n+\frac{1}{2} +\frac{\omega_c
  \Omega^2}{2\widetilde{\omega}^3} \Big(\frac{X}{\ell} \Big)^2\Big] \nn \\
&=& \hbar \widetilde{\omega}\Big[ n+\frac{1}{2} +\frac{\Omega^2}{2\omega_c^2} 
 \Big(\frac{\tilde{X}}{\tilde{\ell}} \Big)^2\Big] \equiv
 \tilde{E}_n(\tilde{X}),
\label{eq:Enpar} \ee
where $\tilde{\ell}^2=\hbar/(m^* \widetilde{\omega}) =(\omega_c/
\widetilde{\omega})\ell^2$,  and 
\be \varphi_{n,X}(x)= \frac{1}{\sqrt{\tilde{\ell}}} \,
u_n\Big(\frac{x-\tilde{X}}{\tilde{\ell}}\Big). \label{eq:phipar} \ee
Since $\tilde{\ell}<\ell$, the parabolic confinement leads, apart from a shift
of their center coordinates, to a reduced width of the wavefunctions.  
A sketch of the model potential, the energy bands, and the squared energy
eigenfunctions is given in the lower part of
Fig.~\ref{fig:parab-wf}. The dotted lines indicate the extent of the
eigenfunctions, which is constant within each Landau band and
increases with the quantum number $n$ of the Landau band. 
The upper part of  Fig.~\ref{fig:parab-wf} shows, for $k_BT=0.01 \hbar
\omega_c$ and three values of the electrochemical potential $\mu^*=E_F$,
the electron and the current density, calculated
 according to  Eqs.~(\ref{eq:dens-fnX}) and
(\ref{eq:curdens-fnX}) with $f_{n,X}=f([E_n(X)-\mu^*]/k_BT)$. On the scale of
the figure, the shown results cannot be distinguished from those calculated
for $T=0$ and the same Fermi energies.

Transforming the 
$X$-integral into an integral over $\tilde{X}$ introduces a pre-factor
$(\widetilde{\omega}/\omega_c)^2$, which increases the density of
effective center coordinates $\tilde{X}$ and thereby of  eigenstates
in each Landau band. Referring the effective filling factor to this enhanced
density of states, $\nu_{\rm eff}(x)=2 \pi \ell^2
(\omega_c/\widetilde{\omega})^2n(x)$, 
leads to the results shown in Fig.~\ref{fig:parab-wf}, with plateau values
equal to integer multiples of the spin-degeneracy $g_s$ in regions where the
Fermi energy is well between two adjacent Landau bands. Note that in
these regions the usually defined filling factor
$\nu(x)=(\widetilde{\omega}/\omega_c)^2 \nu_{\rm eff}(x) > g_s(n_{\rm
  occ} +1)$ is larger than the number of fully occupied
Landau bands with  $n\leq n_{\rm occ}$. 

 The plateaus of the density profile are
separated by broadened steps, which reflect the structure of the
wavefunction of the highest, partly
occupied band. Clearly the width of the density profile increases with
increasing $E_F$, and, due the symmetry of the considered potential, the
profiles are even functions of position. The current density $j_y(x)$, which
is an odd function of position, is, similar to the density, the sum of the
contributions of all (partly) occupied bands, and is determined by the current  
$e\Omega^2 \tilde{X}/\omega_c$ carried by the state $|n,X\rangle$, and
by its occupation. Of course, the total current in the considered
thermal equilibrium state vanishes. 

\begin{figure}[h]
 \includegraphics[width=0.9\linewidth]{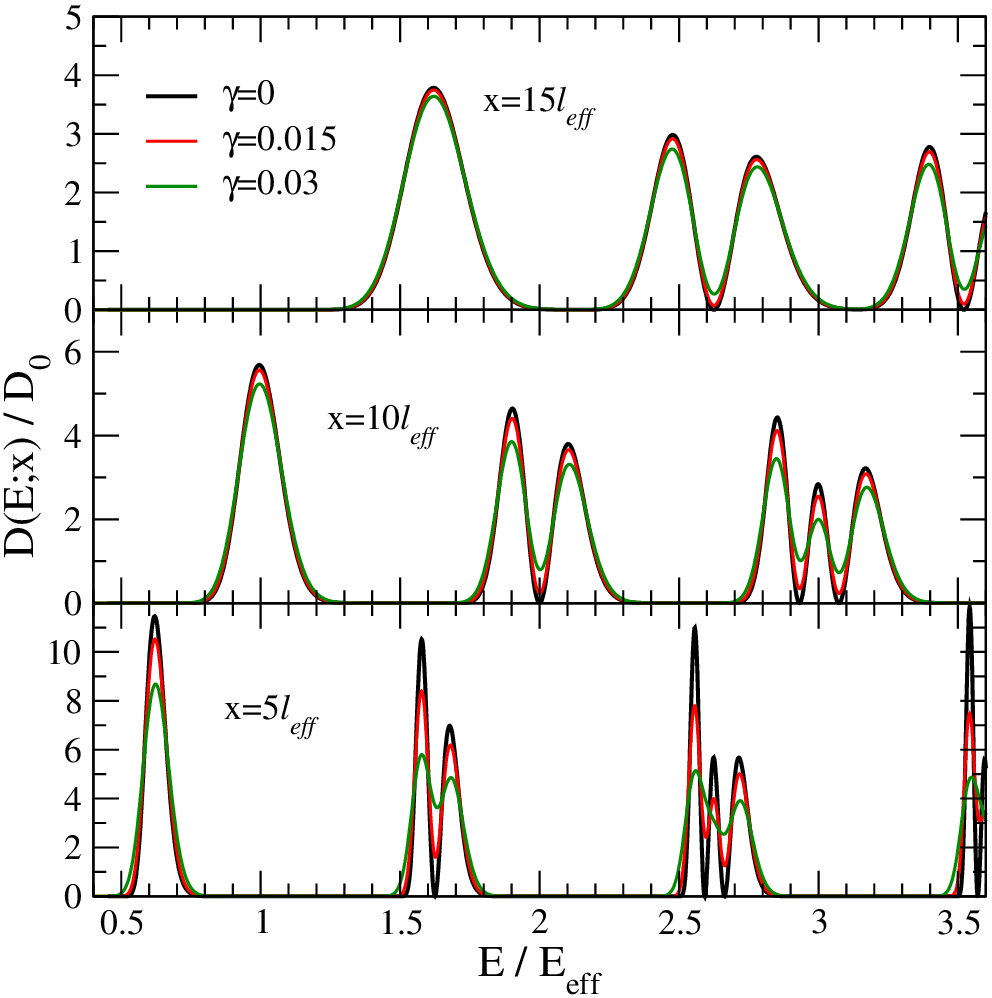}
\caption{\label{fig:parab-ldos} (color online) LDOS $D(E;x)$ for the parabolic
  confinement model with $\Omega/\omega_c=0.1$ at three positions, $x/
  \tilde{\ell}=5,\, 10,\, 15$. The black lines are without collision
  broadening, $\gamma=0$, Eq.~(\ref{eq:ldospar}). The colored lines are with
  collision broadening according to Eqs.~(\ref{eq:plin-cb}) and
  (\ref{eq:Agauss})   for   $\gamma \equiv \Gamma/E_{\rm eff}= 0.015$ and
  $0.03$. Energy in units of $E_{\rm eff}=\hbar \widetilde{\omega} $, position
  in units of $\ell_{\rm eff}=\tilde{\ell}$,
  $D_0=m^*/(\pi \hbar^2)$, and  $g_s=2$. 
}
\end{figure}

The calculation of the LDOS from Eq.~(\ref{eq:ldos-expl}) is also
straightforward, with $\tilde{\xi}=x/\tilde{\ell}$ and $
\tilde{\xi}_n^{\pm}(E)= \tilde{X}_n^{\pm}(E)/\tilde{\ell} $
 we find
\be D(E;x)=\frac{g_s}{2\pi \ell^2} \frac{\widetilde{\omega}^2}{\Omega^2}
\sum_{n, \pm}\frac{u_n^2\big(\tilde{\xi}-\tilde{\xi}_n^{\pm}(E)\big)}
{\hbar \widetilde{\omega} \,
  |\tilde{\xi}_n^{\pm}(E)|}\,\theta_n(E), \label{eq:ldospar} \ee 
where $\theta_n(E)=\theta\big(2E-\hbar \widetilde{\omega}(2n+1)\big)$ and
$ \tilde{X}_n^{\pm}(E)/\tilde{\ell}=\pm (\omega_c/\Omega)\sqrt{2E/(\hbar
  \widetilde{\omega})-(2n+1)}$. Of course $D(E;-x)=D(E;x)$ holds. 
The LDOS is shown for three different positions
$x$ by the black lines in Fig.~\ref{fig:parab-ldos}. Since for increasing
$|x|$ the contribution to the LDOS come from wavefunctions with increasing
$|X|$, i.e.,  increasing values of energy dispersion $|dE_n(X)/dX|$, the
contributions of the individual bands become broader and the gaps between
these contributions become smaller, as can already be seen from
Fig.~\ref{fig:parab-wf}.  The gap between the contributions due to the
lowest bands vanishes for $x/ \tilde{\ell} \gtrsim 20$, i.e. if 
 $\tilde{\ell} |dE_n(X)/dX|/\hbar \widetilde{\omega} \gtrsim 0.2$
holds for the relevant $X$-values. This condition is similar to that found in
the linear-potential model.

Due to the vanishing energy dispersion in the center, $[dE_n(X)/dX](0)=0$, the
contributions of all bands to $D(E;x)$ become very narrow and
$\delta$-function-like for small values of $x$. With increasing distance from
the center, the contributions become wider and clearly reflect the structure
of the corresponding wavefunctions, notably their zeroes. 
\begin{figure}[h]
 \includegraphics[width=0.9\linewidth]{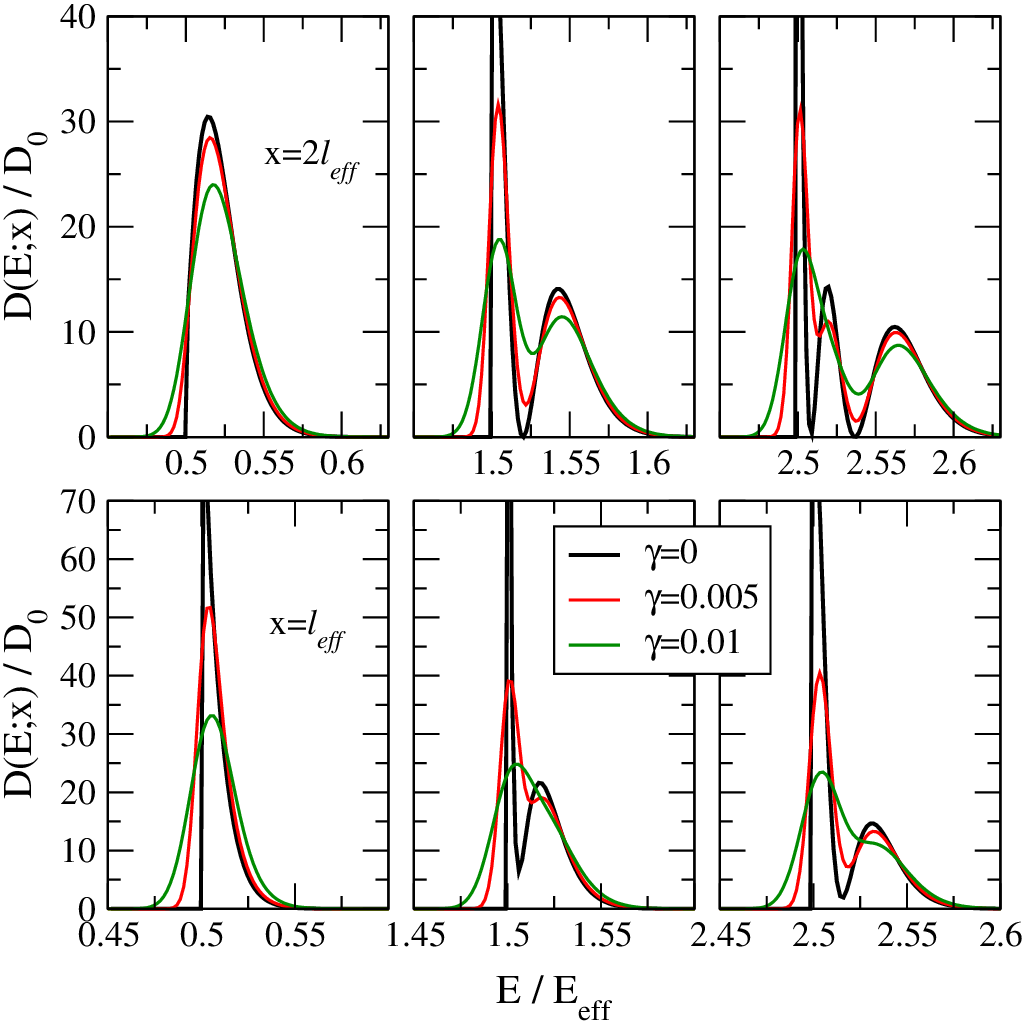}
\caption{\label{fig:parab-smallx} (color online) As in
  Fig.~\ref{fig:parab-ldos}, but for smaller $x$ values and smaller
  damping. The lower three plots are for $x=\tilde{\ell}$, the upper three for
  $x=2\tilde{\ell}$. The gap regions are omitted.
}
\end{figure}
Apparently in the energy range shown in Fig.~\ref{fig:parab-smallx}
all the zeroes of the wavefunctions are
reflected in the LDOS for $x=2\tilde{\ell}$, but not yet for $x=\tilde{\ell}$.

Figures~\ref{fig:parab-ldos} and \ref{fig:parab-smallx}
also show the effect of collision broadening on
$D(E;x)$, which is most important for small $|x|$ values, where it completely
washes out the internal structure of the individual contributions. For larger
values of $|x|$ the collision broadening levels off the maxima and smears out
the zeroes of the individual structures, and at large values, when these
structures become broad, the collision broadening becomes relatively
unimportant.

\section{Model for  incompressible
  stripes \label{sec:is-model} }

In the screening theory of the IQHE in narrow Hall bars
\cite{Guven03:115327,Siddiki04:195335,Gerhardts08:378} incompressible
stripes (ISs)  play an important role, which separate neighboring compressible
regions, in which adjacent Landau levels are pinned to the Fermi energy, since 
these ISs offer the possibility of dissipationless current flow through an
otherwise dissipative Hall bar. To understand the width and the
separation of the QH plateaus, i.e. of the $B$-intervals in which the
resistance quantization occurs, it is important to understand the
conditions under which an IS can carry a dissipation-free current.
The calculations \cite{Guven03:115327,Siddiki04:195335}
were based on a local model for the conductivity tensor $\hat{\sigma}_{\rm
  loc}(x)$, which was obtained from the density-dependent conductivity tensor
$\hat{\sigma}(n_{\rm el})$ of a homogeneous 2DES of density $n_{\rm el}$ by
replacing this density by the local density $n(x)$ of the inhomogeneous
system,   $\hat{\sigma}_{\rm  loc}(x)= \hat{\sigma}\big(n(x)\big) $.
On ISs of finite width with constant integer filling factor the components of
this  $\hat{\sigma}_{\rm loc}(x)$ have the quantized values, and
dissipationless transport is obtained, if the current is restricted to these
ISs.  It
has been argued \cite{Siddiki04:195335,Gerhardts08:378}
that the width of an IS must be sufficiently large,
e.g., more than several times the spatial extent of typical
wavefunctions near the edges of the IS, since otherwise 
wavefunctions from opposite sides of the IS would overlap and lead to
quasi-elastic scattering across the IS. Such QUILLS processes
\cite{Eaves86:346} would lead to dissipation, so that too narrow
stripes between neighboring compressible regions cannot support the
resistance quantization.

A suitable quantity containing quantitative information about the
ability of an IS to carry dissipationless current should be the LDOS
in the IS. To calculate this LDOS, we have to model the IS with some
care. Even if the potential within the IS might be well approximated by
a linear position dependence, at the interesting energies around the
Fermi energy (i.e. the electrochemical potential) there exist nearby
states of the compressible regions, which may contribute to the LDOS
when its gap near the Fermi energy becomes small. 

\begin{figure}[h]
 \includegraphics[width=0.9\linewidth]{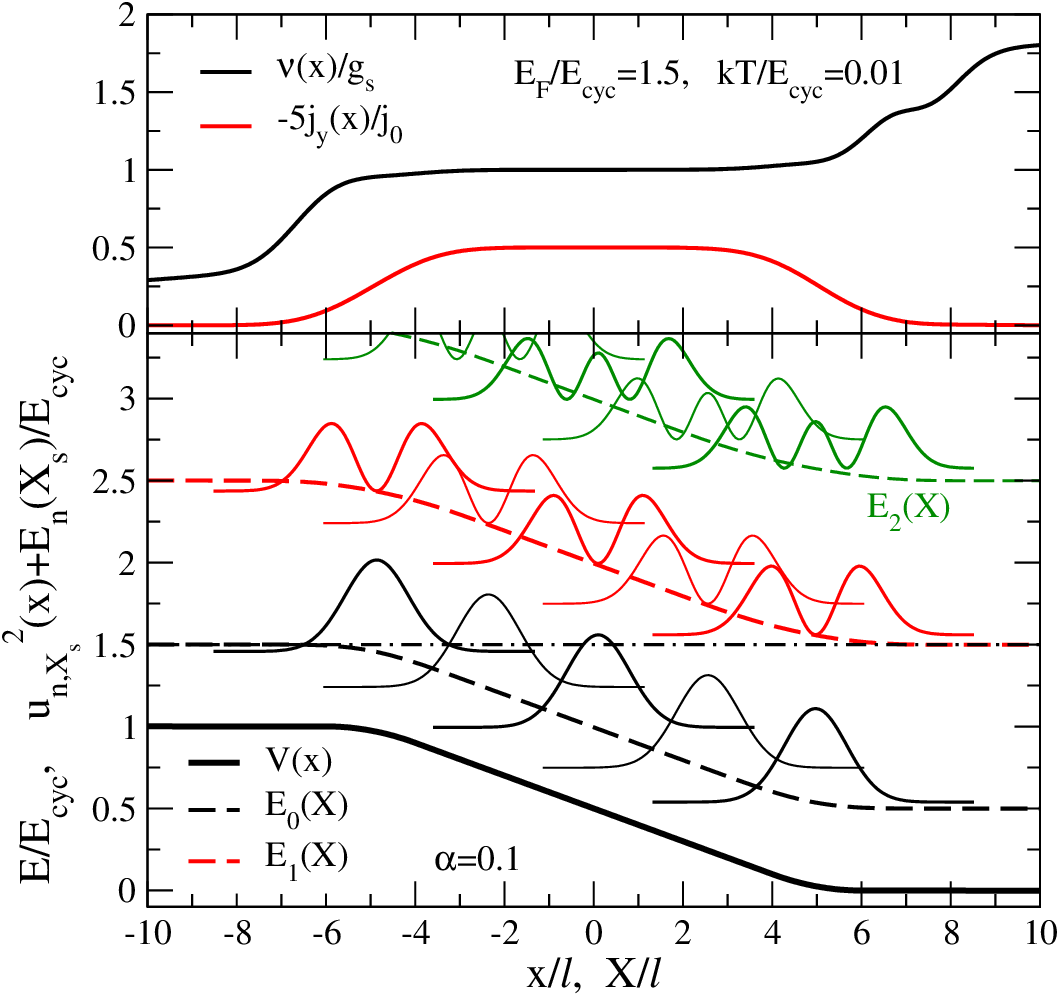}
\caption{\label{fig:is-enwf} (color online) Upper part: filling factor $\nu(x)$
  and current density $j_y(x)$, in units of $j_0=(g_s/2 \pi)\, e \omega_c/
  \ell$; lower part: corresponding
  potential $V(x)$,   energy   eigenvalues $E_n(X)$ and squared
  eigenfunctions  at center coordinates $X_s/\ell=0,\, \pm
  2.46, \, \pm 4.92$, shifted upwards by 
   $E_n(X_s)/\hbar \omega_c$.  } 
\end{figure}
To get an idea how the LDOS changes with the width of an IS, we
consider as a crude model the sum $V(x)=V_S(x)+V_T(x)$ 
of a smoothened step potential
\be  \label{eq:ismodel}
V_S(x)=\hbar \omega_c \cdot \left\{ \begin{array}{cc} 1, & \xi < -\xi_0,\\
  1-\kappa (\xi+\xi_0)^2, & -\xi_0 <\xi < -\xi_+ ,\\
 \frac{1}{2} -\alpha \xi , & -\xi_+ <\xi < \xi_+ ,\\
 \kappa (\xi-\xi_0)^2, &  \xi_+ < \xi < \xi_0,\\
0, & \xi_0 < \xi , \end{array} \right.\ee
and a weak linear potential $V_T(x)=0.1 k_BT \xi$,
with $\xi=x/\ell$. We take $\xi_0=(1-\gamma)/\alpha$, $\xi_+=\gamma/\alpha$,
and $\kappa=\alpha^2/(2-4\gamma)$, with $0<\gamma \leq 1/2$, so that the
fraction 
$2\gamma$ of the step height $\hbar \omega_c$ is bridged by the linear part of
$V_S(x)$ and the total width of the stripe is $2\xi_0$. In the following we
take $\gamma=0.4$ if we want to avoid sharp 
kinks in the potential, or $\gamma=0.5$, if we want to avoid the arbitrarily
introduced smoothening by parabolic potential regions.

 The weak linear term $V_T(x)$
is added to allow for a variation of the electron density in the regions
$|x/\ell|> \xi_0$. The idea is to simulate the situation in compressible
regions, where self-consistent screening leads to pinning of Landau levels to
the Fermi energy, accompanied by a  variation of the  effective potential
$V(x)$ of the order of $k_BT$.

\subsection{Thermal equilibrium}

First we consider the system without imposed current in thermal equilibrium
with constant electrochemical potential $\mu^*=E_F=1.5\hbar \omega_c$, so
that for $X\ll - \ell$ the lowest energy band $E_0(X)$ approaches $E_F$ from
below, and for $X\gg \ell$ the second band $E_1(X)$ approaches $E_F$ from above.
Numerical solution of the eigenvalue problem for this potential $V(x)$ with
$\alpha=0.1$ yields the energy bands and the electron and current densities
presented in Fig.~\ref{fig:is-enwf}. The width ($\sim \ell/\alpha =10 \ell$) of
the IS is large enough to allow for an inner region with constant electron and
current densities, similar to the gap regions of Fig.~\ref{fig:plin-wf-dens},
which shows results for a linear potential with the same slope
$dV/dx=eE_x=-0.1\hbar \omega_c/\ell$.
\begin{figure}[h]
 \includegraphics[width=0.9\linewidth]{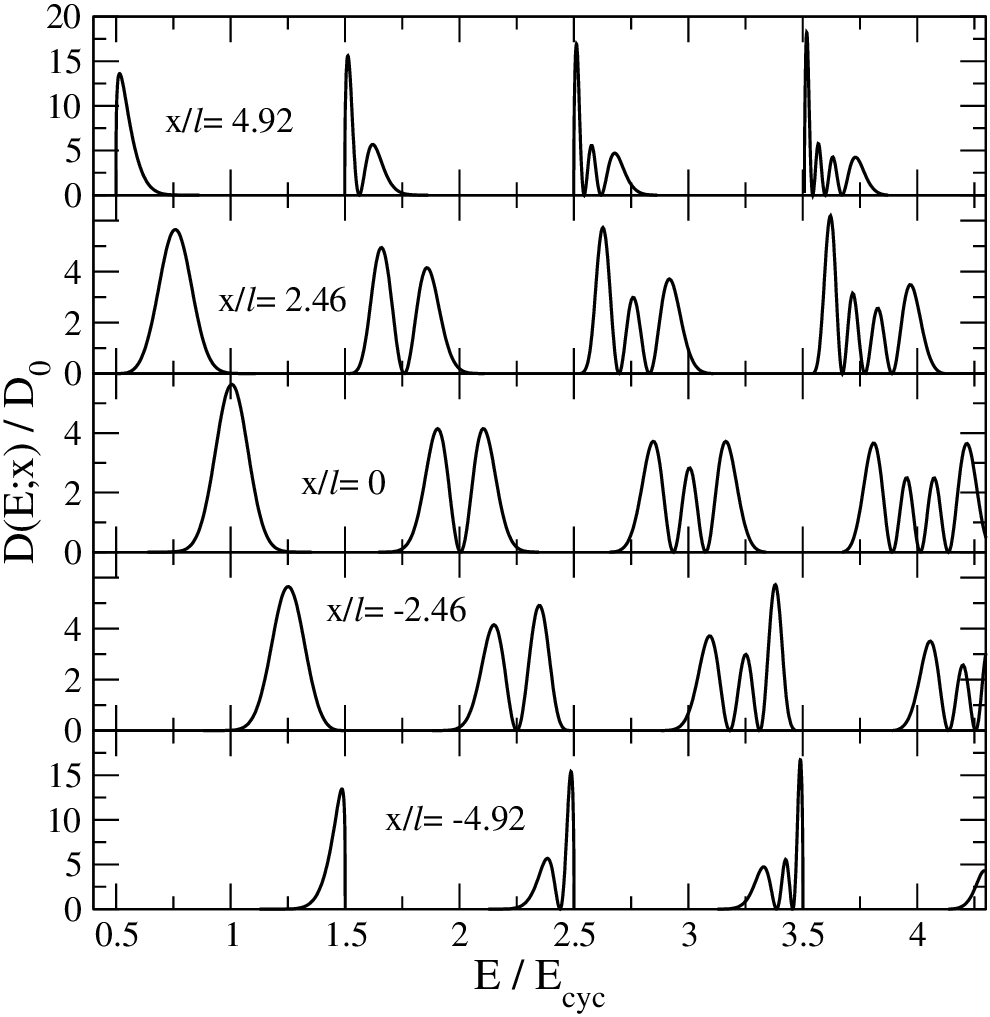}
\caption{\label{fig:isldos} Energy dependence of the LDOS for the five indicated
  positions $x$. The broadened step potential is given by
  Eq.~(\ref{eq:ismodel}) with $\alpha=0.1$ (and $\gamma=0.4$, see text). } 
\end{figure}
 The corresponding LDOS is sketched in Fig.~\ref{fig:isldos}
for five characteristic values of the position $x$. In the center of the IS,
at $x=0$, one finds the same LDOS as for the linear-potential model with the
same electric field, see the lower panel of Fig.~\ref{fig:plin-ldos}.
As the position $x$ moves towards the edges of the IS and leaves the regime of
linear potential, the individual
contributions of the different bands become asymmetric and narrower, since the
magnitudes of the slopes $|dE_n(X)/dX|$ become smaller.

 For positive $x$-values
the curvatures of potential and energy bands become positive, and the
low-energy parts of the individual contributions are enhanced, while the
high-energy parts are reduced, just as we found for the parabolic
confinement potential in Fig.~\ref{fig:parab-ldos}.
 For negative $x$-values the curvatures become negative
and the asymmetry of the individual contributions is inverted, with reduced
low-energy and enhanced high-energy parts. 
\begin{figure}[h]
 \includegraphics[width=0.9\linewidth]{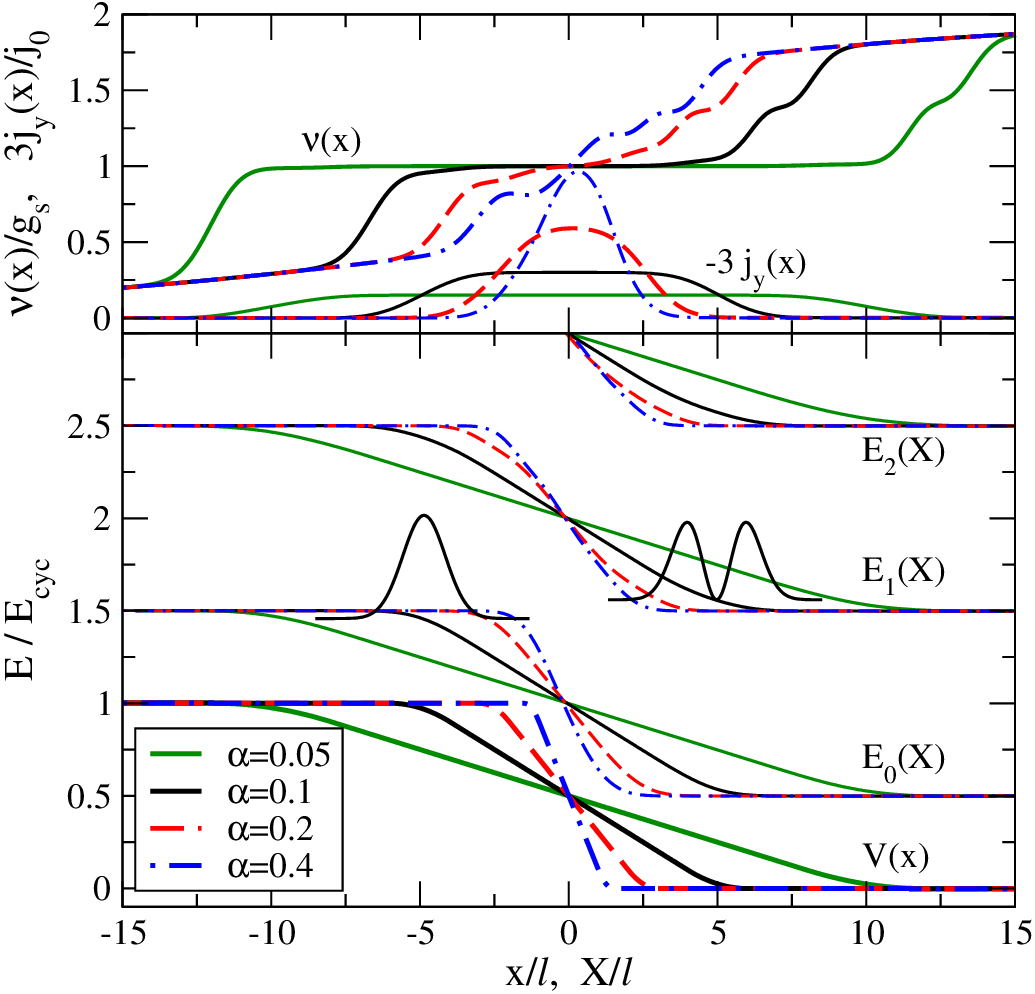}
\caption{\label{fig:is-4alf} (color online) Lower part: Potential
  $V(x)=V_S(x)+V_T(x)$ and corresponding energy bands $E_n(X)$ for $\alpha$
 parameters  as indicated ($\gamma=0.4$). For
  $\alpha=0.1$ shifted squares of energy eigenfunctions are indicated. Upper
  part: corresponding filling factors and current densities, $E_F=1.5
  E_{\rm cyc}$,  $k_BT=0.01 E_{\rm cyc}$, $E_{\rm cyc}=\hbar \omega_c$. } 
\end{figure}
For positions close to the high-energy edge of the IS ($x/\ell \approx -4.9$)
there are many nearby states with lower energy, but nearly no states with
slightly higher energies. As a consequence, the individual band-contributions
to the LDOS for such $x$-values show a sharp high-energy cutoff.
Similarly, near the low-energy edge of the IS ($x/\ell \approx 4.9$) we find
low-energy cutoffs. For positions outside the IS ($|x/\ell|>\xi_0=6$) the
individual band-contributions to the LDOS are extremely narrow and
$\delta$-function-like, very similar to the bare Landau DOS.

Since for the screening theory of the IQHE the existence of incompressible
stripes of finite width is crucial, we present in Fig.~\ref{fig:is-4alf} for
potential steps of different steepness the resulting energy bands and the
electron density and current density profiles. Density plateaus with integer
filling factor $\nu(x)=g_s$ are obtained for $\alpha \lesssim 0.1$.
For $\alpha \gtrsim 0.2$ the potential increase is so steep
that it does not lead to a stripe of constant filling factor. Then it makes no
longer sense to address the step region between the flat parts of the
potential as incompressible stripe. If we assume that the critical
steepness is close to $\alpha=0.15$,  ISs do not exist if the width
of the potential steps is not larger than  $\sim 7\ell$, which is a little
larger than the extent of the low-energy wavefunctions.
\begin{figure}[h]
 \includegraphics[width=0.9\linewidth]{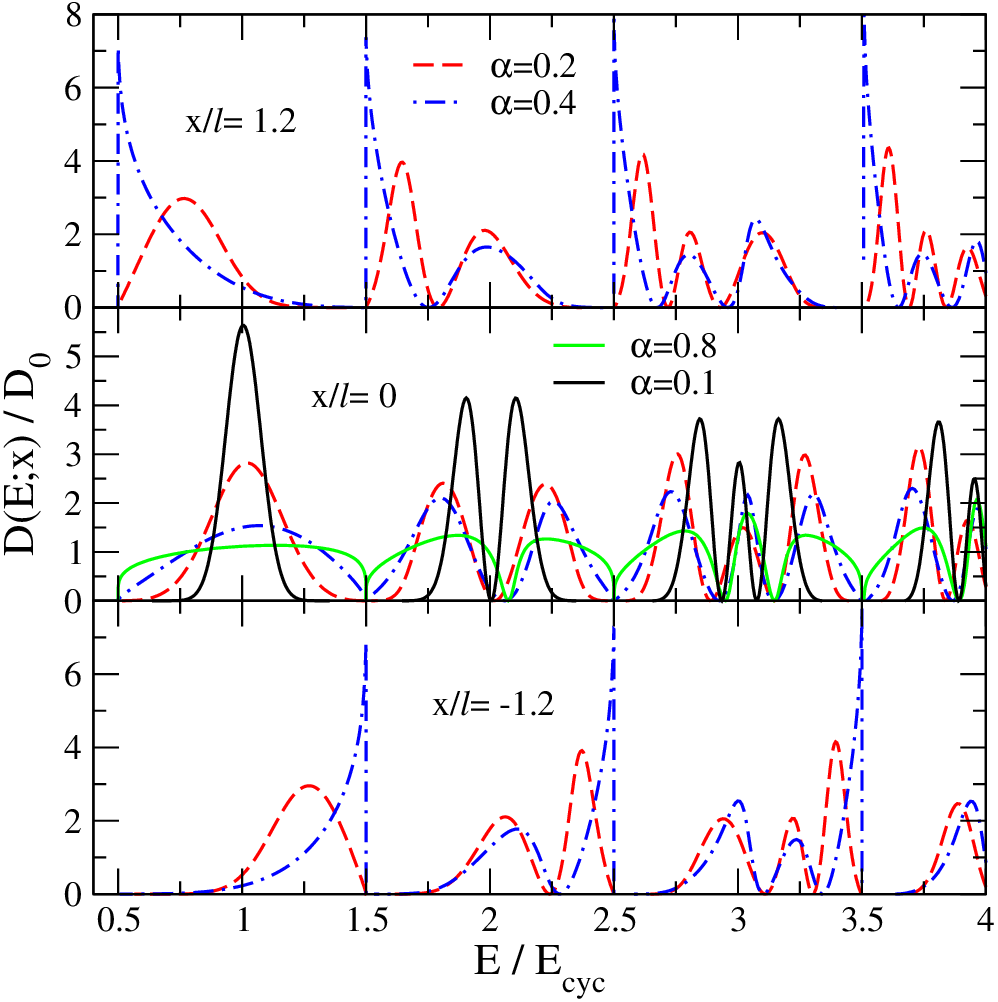}
\caption{\label{fig:isldos3x} (color online)  LDOS at three positions for
  several steepness parameters $\alpha$.} 
\end{figure}
 Figure~\ref{fig:isldos3x} demonstrates how the LDOS
behaves in this limit. The positions $|x/\ell|=1.2$ are for $\alpha=0.2$ well
inside the stripe in the region of weak curvature, and one sees a similar
behavior of the LDOS as  for $|x/\ell|=2.46$ in Fig.~\ref{fig:isldos}. For
$\alpha=0.4$, on the other hand,  $|x/\ell|=1.2$ is at the edge of the stripe
and the situation similar to that for  $|x/\ell|=4.92$ in Fig.~\ref{fig:isldos}.
The situation in the center of the stripe, at $x=0$, is not so easy to
interpret. For $\alpha=0.2$ the contributions of the individual bands to the
LDOS are broader than  for $\alpha=0.1$, as indicated in
Fig.~\ref{fig:isldos3x}, but they are still separated by well developed
gaps, although according to  Fig.~\ref{fig:is-4alf} no IS exists.
 If we increase $\alpha $ further, the gaps shrink, but the LDOS vanishes
at the energies $E=\hbar \omega_c(n+1/2)$, even if the potential step becomes
very narrow. The reason is simple: slightly below $E=1.5\,\hbar \omega_c$
there 
are nearby states in the band $E_0(X)$ and slightly above there are nearby
states of the band $E_1(X)$. But as the energy approaches  $E=1.5 \hbar
\omega_c$, the center coordinates of these states move away from $X=0$ and the
value of their wavefunctions at $x=0$ becomes exponentially small.

This behavior of the LDOS in the center of the IS makes the definition of an
overlap of the contribution of adjacent Landau levels as a criterion for the
vanishing of the gap in the thermal equilibrium situation 
useless. Things change, however, if we consider a situation with imposed
current.

\subsection{Imposed Hall current}
 
We now consider an externally imposed current along the system and assume
(nearly) perfect screening. 
The current is
accompanied by a Hall potential $V_H(x)$, which as a consequence of screening
is constant in the compressible regions  and therefore must
drop over the region of the potential step. Self-consistent screening
calculations    \cite{Guven03:115327} show that the width of the incompressible
stripes is changed by the applied current. It becomes larger, if applied and
intrinsic current have the same direction, and the width becomes smaller, if
applied and intrinsic currents have opposite directions. 
\begin{figure}[h]
 \includegraphics[width=0.9\linewidth]{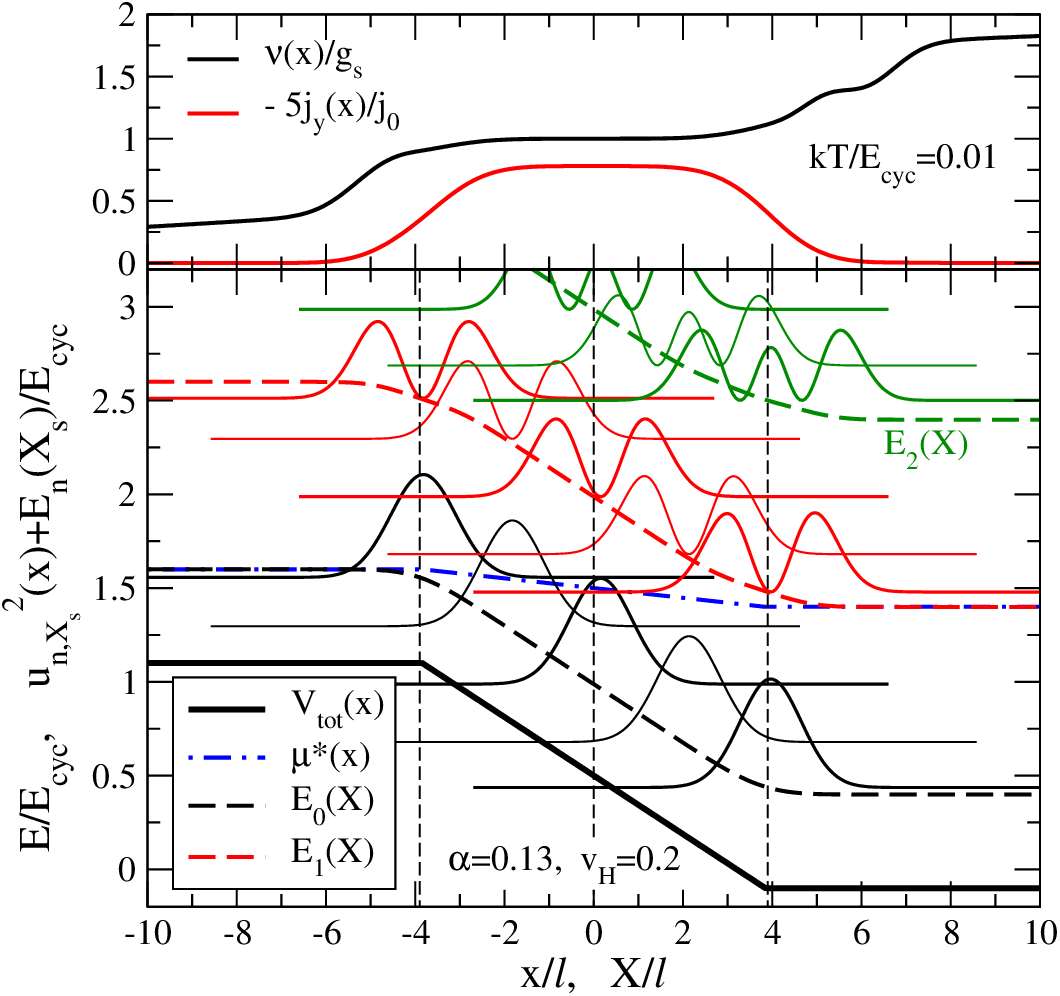}
\caption{\label{fig:is-enwf-Hall} (color online) Lower part: Total potential,
  corresponding energy bands, electrochemical potential, and squared and
  shifted energy eigenfunctions for $X_s/\ell=0, \, \pm 1.98, \, \pm 3.90$.
 Upper part: corresponding filling factors and current densities.
 Parameters    $\alpha=0.13$, $v_H=V_H^0/\hbar \omega_c=0.2$, see text.  } 
\end{figure}
Here we will not
consider these details and make the simplifying assumption that the total
potential $V_{\rm tot}(x)=V(x)+V_H(x)$ and the electrochemical potential
$\mu^*(x)$ are constant in the compressible regions $|x/\ell|>\xi_0$ and vary
linearly in the stripe region, i.e., we put in Eq.~(\ref{eq:ismodel}) 
 $\gamma=0.5$ and $\xi_0=\xi_+=1/(2\alpha)$, and thus suppress the quadratic
 region. To be specific, we take $V_H(x)=V_H^0\, F(x/\ell;\xi_0)$ with
\be F(\xi,\xi_0)=\left\{\ba{cc} \frac{1}{2}, & \xi\leq -\xi_0,\\[0.1cm]
  -\alpha \xi, & |\xi|< \xi_0,\\[0.1cm] - \frac{1}{2}, & \xi \geq \xi_0, \ea
\right. \label{eq:linstep} \ee
as Hall potential, $V_{\rm tot}(x)=V_H(x)+\hbar \omega_c[1/2+F(x/\ell;\xi_0)]$
as total potential, and $\mu^*(x)=1.5\hbar \omega_c+V_H(x)$ as electrochemical
potential. Numerical results for energy bands, wavefunctions, electron and
current density are presented in Fig.~\ref{fig:is-enwf-Hall} for $\alpha=0.13$
and $V_H^0=0.2 \hbar \omega_c$. Although the potential has sharp kinks near
$x=\pm 3.9 \ell$, the energy bands $E_n(X)$ are smooth and the curvatures near
$X=\pm 3.9 \ell$ become smaller with increasing $n$. In the linear regime near
$x=0$ the centers of the wavefunctions $\varphi_{n,X}(x)$ are shifted from $X$
to larger values, as expected from Eq.~(\ref{eq:fitilde}). Near $x=3.9\ell$,
where the potential kink can be considered as limit of a positive curvature,
the wavefunctions are somewhat narrower than near $x=0$. This is immediately
understood from Eq.~(\ref{eq:phipar}) and a parabolic approximation of the
potential near the kink. Similarly, near   $x=-3.9\ell$, where the potential
kink corresponds to a negative curvature, the wavefunction are somewhat wider
than near $x=0$.
\begin{figure}[t]
 \includegraphics[width=0.9\linewidth]{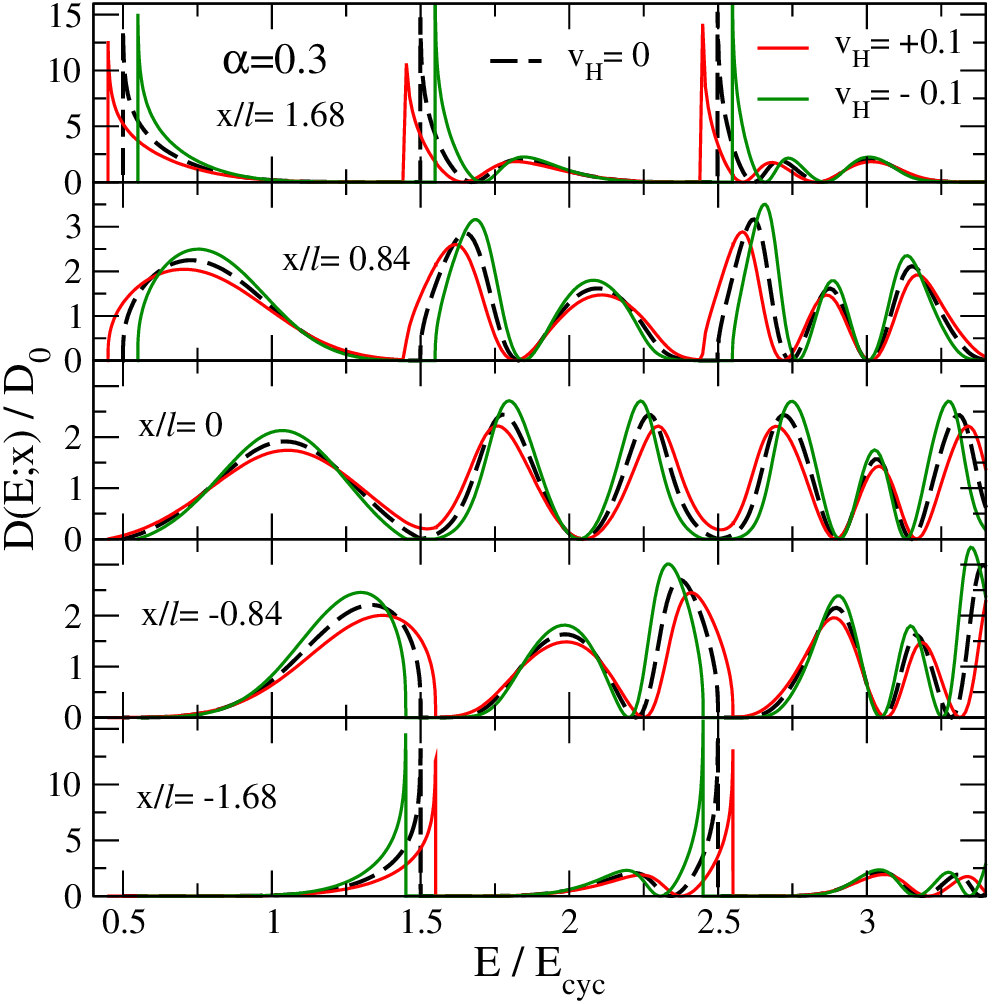}
\caption{\label{fig:is-ldos-Hall} (color online) LDOS $D(E;x)$ for a steep
  potential step, $\alpha=0.3$, with and without externally imposed current, 
  for five positions $x$ as indicated. Strength and direction of the imposed
  currents are characterized by the corresponding Hall potentials
  $v_H=V_H^0/\hbar\omega_c$ with values $0, \, \pm0.1$ as indicated.  } 
\end{figure}

The most important consequence of the imposed current is that the
corresponding Hall potential leads to an energetic overlap of the high-energy
edge of the lowest energy band $E_0(X)$ and the low-energy edge of the next
band $E_1(X)$. Thus the situation near $x=0$ is similar to that in the
linear-potential model, and the individual band-contributions to the LDOS will
overlap, if the region of the potential step will become to narrow.
If the external current is applied in the opposite direction to that of the
intrinsic current, the sign of $V_H(x)$ will change and, instead of an overlap
of the bands, a finite energy gap between the bands will result. This will
lead to energy gaps in the LDOS at $x=0$, which will remain even if the width
of the potential step will become small.

These results for the LDOS are illustrated in Fig.~\ref{fig:is-ldos-Hall},
where we consider a steep potential step, $\alpha=0.3$, which does not allow
for a IS with constant density in the step region.   
Without imposed current, $v_H=0$, the LDOS at $x=0$ has no gaps, but has very
small values in the middle $E_{n,n+1}=[E_n(0)+E_{n+1}(0)]/2$ between the band
energies $E_n(X)$ at $X=0$, $D(E_{n,n+1};0)\approx 0$. If a current in the
direction of the intrinsic current is imposed, $v_H=0.1$, these values
increase drastically and a considerable overlap of the contribution due to
adjacent bands is observed. If the current is imposed in the opposite
direction, $v_H=-0.1$, well developed  gaps occur around the energies
$E_{n,n+1}$, in which $D(E;0)$ vanishes. On the other hand, the density
profile, which is not shown, changes only very little due to the applied
current and remains without 
any flat part in the step region for both directions of the imposed current,
similar to the profiles shown in the upper part of  Fig.~\ref{fig:is-4alf} for
$\alpha \geq 0.2$.

\begin{figure}[t]
 \includegraphics[width=0.9\linewidth]{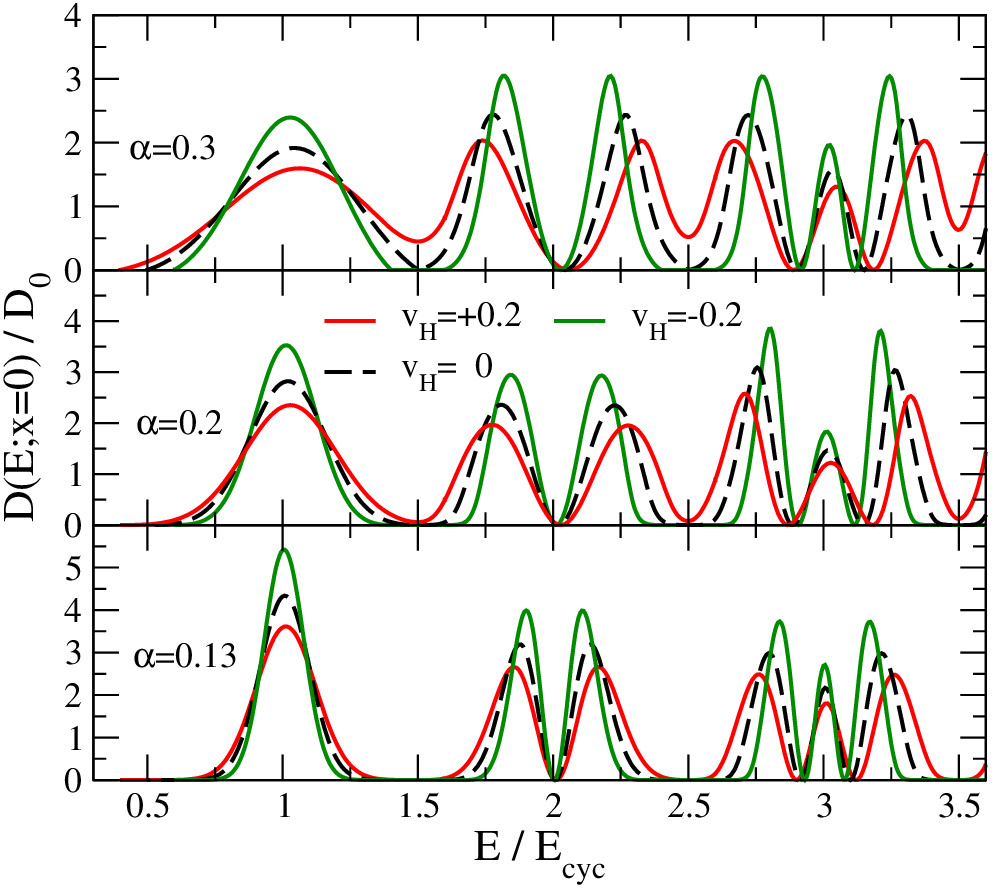}
\caption{\label{fig:is-ldos-x0-vh} (color online) LDOS $D(E;x=0)$ for
  three value of the steepness, $\alpha=0.13, \, 0.2,\,0.3$ and three
  values of the Hall voltage, $v_H=V_H^0/\hbar\omega_c=0,\, \pm 1$, as
  indicated.  } 
\end{figure}

Figure~\ref{fig:is-ldos-x0-vh} demonstrates how gaps and overlap,
respectively, of the LDOS $D(E;x)$ in the center of the potential
step, $x=0$,  depend on the steepness of the potential and on the Hall
potential.
\begin{figure}[h]
 \includegraphics[width=0.85\linewidth]{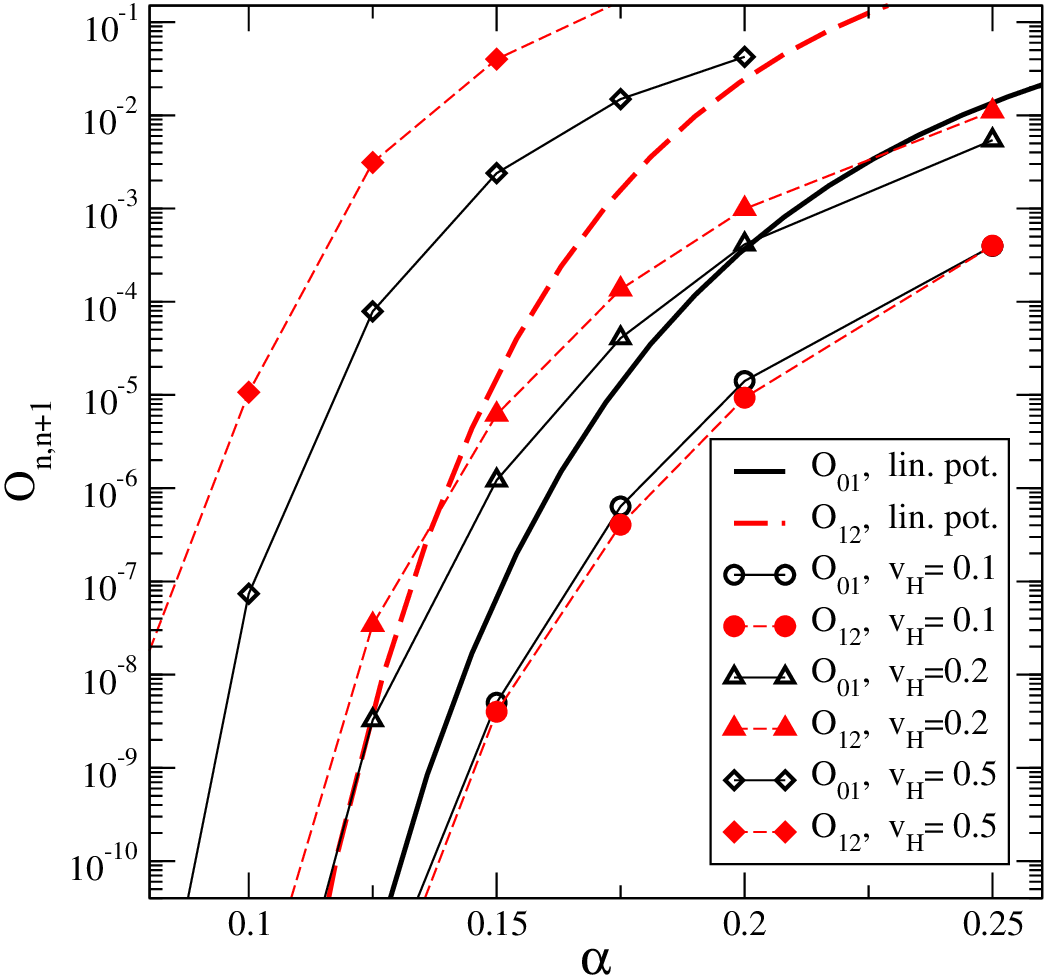}
\caption{\label{fig:is-overlp} (color online) Overlaps $O_{01}$ and
  $O_{12}$ as functions of the potential steepness $\alpha$ for
  different value of the reduced Hall voltage $v_h=V_H^0/\hbar
  \omega_c$, as indicated by the symbols. The lines between the
  symbols are guides to the eye. The heavy lines indicate the
  corresponding results for the linear-potential model, shown in
  Fig.~\ref{fig:overlp} for $\eta \equiv \alpha$.  } 
\end{figure}
For $V_H^0<0$ we observe well developed gaps, even if the potential
step is so steep, that no IS exists ( see Fig.~\ref{fig:is-4alf}). For 
$V_H^0>0$, i.e. imposed and intrinsic current in the same direction,
we have a situation as in the linear-potential model, and we can
consider the overlap as a function of the potential steepness, as in
Fig.~\ref{fig:overlp}. Generalizing Eq.~(\ref{eq:overlap}) by
\be O_{n,n+1}=D_n(E_{n,n+1};0) D_{n+1}(E_{n,n+1};0)/D_0^2,
\label{eq:verlap-is} \ee
where the $D_n(E;x)$ are the individual band contributions to
$D(E;x)=\sum_n D_n(E;x)$ defined in Eq.~(\ref{eq:ldos-expl}). 
Results are shown in Fig.~\ref{fig:is-overlp}. 

\section{Remarks and conclusion \label{sec:conclusion}}

We have simplified and extended previous calculations
of the LDOS of  a Landau quantized 2DES in the presence of
a constant, unidirectional in-plane electric field,
\cite{Kramer04:21,Kramer06:1243}  and we have shown that the
LDOS in principle is a useful concept for further calculations, if the
occupation probability of energy eigenstates depends only on their energy
eigenvalues but not on other conserved quantities. We have also considered the
case of a homogeneous 2DES supporting a homogeneous dissipation-free Hall
current, which can be described by the standard methods of grand-canonical
equilibrium \cite{Oh97:13519} but not in terms of the LDOS. 
For this linear-potential model we have also quantified the overlap of
adjacent band contributions to the LDOS, which leads to a closing of gaps in
its energy-dependence and indicates the onset of quasi-elastic
inter-Landau-level scattering (QUILLS). In realistic situations QUILLS will
lead to a breakdown of the IQHE, e.g., under high externally imposed
currents. To get more than an indication of the onset of such breakdown
effects, one should, however, explicitly consider electron-impurity scattering
under these conditions, which goes far beyond the mere calculation of the LDOS
of the idealized 2DES. Here we have considered only a simple phenomenological 
treatment of collision broadening and mentioned that this is not sufficient
under strong electric fields. Future work on the necessary generalization of the
treatment of collision broadening, even within the frame of the self-consistent
Born approximation, seems desirable.

To get a better understanding of the behavior of the LDOS in other situations
than the simple linear-potential model, we have considered 
a parabolic potential as a model for a laterally confined 2DES, which also
allows analytical calculation of the LDOS, Eq.~(\ref{eq:ldospar}). Whereas
$D(E;x)$ at $x=0$ exhibits 
one-over-square-root singularities of the type $1/\sqrt{E-E_n(0)}$ at the
energies $E>E_n(0)$, for $|x|>0$ there are no singularities but, due to the
increasing electric field strength, the individual band contributions to the
LDOS become asymmetric and broader with increasing $|x|$.

Finally we have calculated the LDOS for incompressible stripes, which are an
essential ingredient of the screening theory of the
IQHE. \cite{Guven03:115327,Siddiki04:195335,Gerhardts08:378} We use a
simplified model of such an IS, which describes the compressible regions (CRs)
next 
to the stripe by nearly constant potentials, and the stripe region in between
by a more or less linear potential. One expects that, without collision
broadening, the LDOS in the CRs approaches the singular
Landau DOS, whereas in the linear-potential regions of incompressible stripes
the situation should be similar to that in the linear-potential model
considered in sect.~\ref{sec:ldos2}. In the transition regions between
nearly constant and linear potential the results should be comparable with
those for the parabolic-potential of sect.~\ref{sec:parabpot}. Concerning the
energy dependence of the LDOS at characteristic positions $x$ we find these
expectations confirmed. However, the closing of energy gaps and the onset of
overlap of contributions from adjacent bands, which we interpreted in the
linear-potential model of sect.~\ref{sec:ldos2}
as indication for the breakdown of the IQHE, are now more subtle and not so
easy to interpret.  In the center of the stripe region at $x=0$, the energy
gaps become smaller as the distance between the CRs decreases and at the
energies above and below the Fermi energy the centers of the energy
eigenfunctions move towards the center of the stripe region. But exactly at
the Fermi energy, which separates the lowest energy bands $E_0(X)$ and
$E_1(X)$, there are no nearby states and $D(E_F;0)\approx 0$ even if the CRs
come so close that, due to the finite extent of the wavefunctions, no genuine
IS with constant local filling factor $\nu(x)=g_s$ in an $x$-interval of finite
width exists. 
The situation becomes clearer, if one imposes an external current on the
system. This leads to a Hall potential in the stripe region, which may
increase or diminish the intrinsic potential variation across the stripe
region. If the imposed current has the same direction as the intrinsic one,
the potential variation increases and around $E_F$ the two lowest bands overlap
energetically. Then the situation is as in the linear-potential model
of sect.~\ref{sec:ldos2}, and the overlap criterion for the breakdown of the
IQHE can be applied. If imposed and intrinsic currents have opposite
directions, the potential variation across the strip region decreases and at
$E_F$ a gap opens between the two lowest bands. Then in $D(E;0)$ an energy gap
of finite width around $E=E_F$ remains, even if the distance between the CRs
becomes so small that, the stripe region between them can no longer support a
dissipationless current, i.e. support the IQHE. Of course, in a real sample
both situations occur simultaneously, since the intrinsic currents in the
stripe regions of opposite sides of the sample have opposite directions.
Within the screening theory of the IQHE one finds that the width of the
incompressible stripes is different in both situations. If imposed and
intrinsic currents have the same direction, the stripe is wider than in the
opposite case, \cite{Guven03:115327,Gerhardts08:378} 
however, to determine the widths of the stripes and to decide whether they can
support the IQHE requires an involved self-consistent calculation. 

In summary, the LDOS is an interesting concept, is easy to evaluate, and can
give some hints on possible scattering effects, such as QUILLS, which may lead
to the breakdown of the IQHE. However, to really understand the IQHE is much
more complicated and requires non-trivial calculations. One should include the
relevant scattering effects, which usually lead to dissipative transport, and
find out, under which conditions they become ineffective and lead to the
peculiar transport phenomena observed in the plateau regime of the IQHE.

\acknowledgments
We thank T. Kramer for drawing our attention to the LDOS concept, and for
fruitful discussions. E. B. Sa\~gol is acknowledged, for pointing out
experimental details and related literature. This work is partially supported
by T\"UBiTAK under grant no:109T083 and by IU-BAP:6970.

\end{document}